\documentclass[journal,twocolumn,10pt,twoside]{IEEEtran}
\usepackage{graphicx,cite,epsfig,amssymb,amsmath}
\usepackage{pifont}
\usepackage{stmaryrd}
\usepackage{bbding}
\usepackage{subfigure}
\usepackage{setspace}
\usepackage{color}
\usepackage{slashbox}
\usepackage{algorithm,algorithmic}
\usepackage{bm}
\usepackage{subfigure}
\allowdisplaybreaks[4]

\begin{document}

\title{Massive Unsourced Random Access over Rician Fading Channels: Design, Analysis, and Optimization}
\author{\normalsize
Feiyan Tian, \textit{Student Member, IEEE}, Xiaoming Chen, \textit{Senior Member, IEEE}, Lei Liu, \textit{Member, IEEE}, \\and Derrick Wing Kwan Ng, \textit{Fellow, IEEE}
\thanks{Feiyan Tian and Xiaoming Chen are with the College of Information Science and Electronic Engineering, Zhejiang University, Hangzhou 310027. Lei Liu is with the School of Information Science, Japan Advanced Institute of Science and Technology, Ishikawa Prefecture, Japan. Derrick Wing Kwan Ng is with the school of Electrical Engineering and Telecommunications University of New South Wales, Sydney, NSW 2052, Australia.}}\maketitle

\begin{abstract}
In this paper, we investigate an unsourced random access scheme for massive machine-type communications (mMTC) in the sixth-generation (6G) wireless networks with sporadic data traffic. Firstly, we establish a general framework for massive unsourced random access based on a two-layer signal coding, i.e., an outer code and an inner code. In particular, considering Rician fading in the scenario of mMTC, we design a novel codeword activity detection algorithm for the inner code of unsourced random access based on the distribution of received signals by exploiting the maximum likelihood (ML) method. Then, we analyze the performance of the proposed codeword activity detection algorithm exploiting Fisher Information Matrix, which facilitates the derivative of the approximated distribution of the estimation error of the codeword activity vector when the number of base station (BS) antennas is sufficiently large. Furthermore, for the outer code, we propose an optimization algorithm to allocate the lengths of message bits and parity check bits, so as to strike a balance between the error probability and the complexity required for outer decoding. Finally, extensive simulation results validate the effectiveness of the proposed detection algorithm and the optimized length allocation scheme compared with an existing detection algorithm and a fixed length allocation scheme.
\end{abstract}

\begin{IEEEkeywords}
6G, mMTC, grant-free, unsourced random access, Rician fading channel.
\end{IEEEkeywords}

\section{Introduction}
With the upcoming Internet-of-Things (IoT) era, the sixth-generation (6G) wireless networks are expected to provide numerous machine-type communication services, e.g., sensing, monitoring, and reporting \cite{UN0,5G1,5G2}. In general, machine-type communications in 6G wireless networks inherit the characteristics of massive connectivity, sporadic traffic, and small payload. It is expected that 6G wireless networks support 10 devices per $m^2$ in 2030. For such massive machine-type communications (mMTC), conventional grant-based random access schemes may lead to exceedingly high access latency and prohibitive signaling overhead. To address these challenging issues, it is necessary to design new random access schemes according to the characteristics of extreme massive connectivity in 6G wireless networks \cite{6G}.

Recently, grant-free random access, as a new random access scheme, has been introduced into mMTC for handling massive access \cite{GF1,GF2}. Unlike grant-based random access, grant-free random access allows active devices to access wireless networks directly without a prior authority such that the access latency can be reduced significantly in the scenarios of massive connectivity and sporadic traffic. Currently, there are two kinds of grant-free random access schemes \cite{GF20}. The first one is sourced random access, in which the base station (BS) has to first know the identities of active devices and then recovers their transmitted data packets. Therefore, the active devices should send unique pilot sequences over the uplink channels for performing device activity detection and channel estimation. Since device activity detection can be regarded as a typical sparse signal recovery problem, numerous compressive sensing (CS)-based approaches have been applied to design sourced random access algorithms \cite{GF3,GF4,GF5}. Generally speaking, the access procedure of sourced random access contains two steps. First, active devices send predetermined pilot sequences to the BS for activity detection and channel estimation at the beginning of a time slot \cite{GF51,GF52}. Then, active devices transmit data signals in the rest of the time slot. Nonetheless, in the context of massive connectivity, to guarantee the accuracy of activity detection and channel estimation, active devices are forced to send long pilot sequences, introducing long end-to-end delay. Even worse, the traffic of mMTC is usually with small payload and thus long pilot sequences occupy most of the packet length, leading to the underutilization of limited system resources.

The other practical scheme is unsourced random access, in which the BS is only interested in the transmitted messages, but is not concerned about the identities of active devices which sent the messages \cite{UN1,UN2,UN8,UN9,UN10}. In this context, sending unique pilots in advance is not necessary for device detection, which reduces the consumption of wireless resources significantly, especially in the case of massive connectivity. Since unsourced random access is unconcerned with the identities of active devices, all devices can share the same codebook. In particular, the message of each active device is mapped to a codeword of the codebook and then the corresponding codeword is transmitted for massive access. Consequently, the task of the BS is changed from device activity detection to codeword activity detection. If any devices wish to mark themselves, they can include their identities into the data information. Note that even if the number of transmitted information bits is small, e.g., tens of bits, the size of the codebook can be huge, resulting in intolerable computational complexity for codeword activity detection. Hence, to address this issue, the idea of segmentation is naturally applied to unsourced random access. Specifically, each segmentation is independently sent based on a small codebook such that the computational complexity for codeword activity detection is reduced. It is worth pointing out that for unsourced random access, the exceedingly long preamble sequences are avoided. Although redundancy will be added in the case of message segment, the amount of redundancy is limited because it is independent of the number of devices. Thus, unsourced random access is more efficient than sourced random access.
For instance, the authors of \cite{UN2} introduced a coded CS scheme for unsourced random access over the adder channel. In this scheme, the message transmission process involves a two-layer coding, i.e., inner code (CS code) and outer code (tree code). Specifically, at the sender, a tree-based outer encoder splits the long messages into sub-blocks and the CS inner encoder is responsible for mapping each sub-block to a codeword of a common codebook. Then at the receiver, the CS inner decoder completes the codeword activity detection from the noisy received signals over all sub-blocks and finally the tree-based outer decoder connects the sub-blocks to recover original messages.
Based on the proposed framework in \cite{UN2}, a non-Bayesian codeword activity detection algorithm was proposed for the inner decoder of massive unsourced random access in \cite{UN4}, which outperforms the Bayesian vector approximate message passing schemes in terms of stability. Meanwhile, the authors of \cite{UN5} redesigned the unsourced random access framework in \cite{UN2}. In particular, the inner decoder and the outer decoder operated cooperatively and passed messages back and forth to take the advantages of the coded CS structure to ensure reliable propagation. Furthermore, the authors of \cite{UN6} proposed a distinct uncoupled CS-based unsourced random access scheme, where the tree-based encoding was cancelled and the tree-based decoder was replaced with a clustering decoder. Then, the sub-blocks recovered by the inner decoder were stitched in the clustering decoder by leveraging the inherent coupling of channel correlations across sub-slots.

Despite various studies have solved many practical problems, a common drawback of previous works is that they only considered unsourced random access over Rayleigh fading channels \cite{UN2,UN4,UN5,UN6,UN11,UN12}.
As a matter of fact, due to a very high connectivity density in mMTC, e.g., $10$ devices per $m^2$ in 6G wireless networks, the access distance may be within hundreds or even tens of meters. Hence, the line-of-sight (LOS) propagation is an indispensable scenario in practical mMTC \cite{Rice1,Rice2}. Unfortunately, Rayleigh fading channels do not include a consistent LOS component. In contrast, Rician fading channel contains both LOS and non-LOS components which is more appropriate and accurate for mMTC. Besides, the channel distribution directly determines the statistical information of the received signals at the BS, which is the key of the statistics based signal processing schemes. In other words, existing results designed for Rayleigh fading channels are not applicable to the Rician fading counterpart systems. Additionally, the recent work \cite{UN13} studied a massive unsourced random access scheme for correlated channels with LOS propagations. This work focused to design a activity detection algorithm, but did not provide a complete system framework or realize the trade-off between the complexity and the probability of error. Therefore, to the best of authors' knowledge, massive unsourced random access over Rician fading channels is still an open issue.

In order to improve the performance of unsourced random access schemes, we aim to provide a universal solution for massive unsourced random access over Rician fading channels in 6G wireless networks. The contributions of this paper are as follows:
\begin{enumerate}

\item Inner coding: Due to a very large connection density for mMTC in the sixth-generation wireless networks, unsourced random access may experience Rician channel fading with a consistent LOS component. The existing works designed the inner coding algorithms assuming Rayleigh channel fading without a consistent LOS component. Intuitively, the direct adoption of the inner coding algorithms based on the Rayleigh channel fading assumption may lead to a severe performance degradation, which has been confirmed by simulations in this paper. To this end, we design an inner coding algorithm based on the Rician channel fading assumption. Moreover, we design a flexible threshold for the decision of active codeword according to the characteristics of Rician fading, which is widely applicable with respect to the fixed threshold in previous relevant works.

\item Outer coding: mMTC usually adopts short-packet transmission schemes due to its small data characteristic. In this context, the size of a data packet is limited, which affects both performance and complexity of the unsourced random access scheme. In order to achieve a balance between performance and complexity, we design an outer coding algorithm by optimizing the length of information bits and check bits in the minimization of the decoding complexity subject to a constraint on the probability of decoding failure in the presence of Rician channel fading. Thus, it is possible to design the outer coding algorithm according to the requirements of performance and complexity.

\end{enumerate}

The rest of this paper is organized as follows: Section II discusses the adopted model of a 6G mMTC system over Rician fading channels. Section III provides a massive unsourced random access framework with two-layer message coding and designs a novel codeword activity detection algorithm for the inner code. Section IV analyzes the performance of the proposed codeword activity detection algorithm by deriving the distribution of estimation error and optimizes the lengths of parity check bits and message bits for the outer code.
Then, extensive simulation results are presented in Section V to validate the effectiveness of the proposed scheme. Finally, Section VI concludes the paper.

\emph{Notations}: We use bold upper (lower) letters to denote matrices (column vectors), $(\cdot)^T$ to denote transpose, $(\cdot)^H$ to denote conjugate transpose, $\mathbb{E}[\cdot]$ to denote expectation, $\mathbb{C}^{a \times b}$ to denote a complex matrix or vector of dimension $a \times b$, $\mathcal{CN}(\textbf{x},\textbf{Y})$ to denote the complex Gaussian distribution of a vector with mean vector $\textbf{x}$ and covariance matrix $\textbf{Y}$, $\exp(\cdot)$ to denote the exponent, $\mathbb{P}(\cdot)$ to denote the probability of an event, $\textmd{diag}(\textbf{x})$ and $\textmd{diag}(\textbf{A})$ to denote the matrix with $\textbf{x}$ as the diagonal and the diagonal vector of matrix $\textbf{A}$, respectively, $\textbf{X}_{:,k}$ to denote the $k$-th column of $\textbf{X}$, decimal($\cdot$) to denote the decimal representation of binary sequence, $\mathbb{R}^a_+$ and $\mathbb{N}^a_+$ to denote the non-negative real and integer vector of dimension $a$, respectively, $\triangledown$ to denote gradient, $d(\cdot)$ to denote the differentiation, $|\cdot|$ and $\textmd{tr}(\cdot)$ to denote the determinant and the trace of a matrix, respectively.

\section{System Model}
We consider a single-cell 6G mMTC system based on the unsourced random access protocol, where a BS equipped with $M$ antennas serves $K_{\textmd{tot}}$ single-antenna IoT user equipments (UEs) over the same time-frequency resources. Note that the connectivity density in 6G wireless networks is generally high, i.e., 10 UEs per $m^2$. However, in a certain coherence time interval, only a small set of $K_a$ $(K_a \ll K_{\textrm{tot}})$ UEs, denoted by $\mathcal{K}_a$, are active due to the sporadic nature of IoT applications \cite{GF3,GF4,GF5}. Each active UE intends to transmit $b$ bits of message to the BS and the collection of messages is represented by $\mathcal{L}=\{\textbf{m}_k \in \mathbb{B}^{b\times 1} : k \in \mathcal{K}_a\}$. It is worth pointing out that due to unsourced random access, all UEs share the same codebook. Each active UE maps its message to a codeword of the codebook and then sends it to the BS. Consequently, the BS does not identify which UEs send the codewords, but only aims to recover the transmitted codewords. If UEs wish to mark themselves, they can include their identities into the data messages.

Suppose that the channel between each UE and the BS experiences block fading with a coherence time interval of length $n$, i.e., the channel maintains a constant in each interval and fades independently over intervals. Considering the dense distribution of UEs in the mMTC system, the uplink channel vector $\textbf{h}_k \in \mathbb{C}^{M \times 1}$ from the $k$-th UE to the BS is modeled as the following Rician distribution:
\begin{equation}\label{channel}
\textbf{h}_k=\sqrt{\frac{\kappa_k}{\kappa_k+1}}\ \textbf{h}_k^L + \sqrt{\frac{1}{\kappa_k+1}}\ \textbf{h}_k^S,\ \forall k,
\end{equation}
where $\kappa_k$ is the Rician factor of the $k$-th UE. $\textbf{h}_k^S \sim \mathcal{CN}(\textbf{0},\textbf{I}_M)$ denotes the scattered or non-LOS component and $\textbf{h}_k^L$ represents the deterministic LOS component. For convenience, we define
\begin{eqnarray}
\textbf{g}_k&=&\sqrt{\frac{\kappa_k}{\kappa_k+1}}\ \textbf{h}_k^L,\ \forall k,\\
\alpha_k&=&\frac{1}{\kappa_k+1},\ \forall k.
\end{eqnarray}
Then, we have
\begin{equation}\label{hDistri}
\textbf{h}_k=\textbf{g}_k + \sqrt{\alpha_k}\ \textbf{h}_k^S\ \ \sim\ \ \mathcal{CN}(\textbf{g}_k,\alpha_k\textbf{I}_M),\ \forall k.
\end{equation}
Therefore, the $t$-th signal sample of the received signal at the $m$-th antenna can be expressed as
\begin{eqnarray}
y_{t,m}=\sum\limits_{k \in \mathcal{K}_a}\sqrt{\beta_k}h_{k,m}c_{k,t}+z_{t,m},\nonumber\\\forall t \in \{1,...,n\},\ m \in \{1,...,M\} ,
\end{eqnarray}
where $\beta_k$ represents the channel large-scale fading coefficient and $h_{k,m}$ is the $m$-th element of $\textbf{h}_k$. $c_{k,t}\in \mathbb{C}$ denotes the $t$-th complex-valued symbol of the codeword sent by UE $k$ and satisfies the power constraint $\sum_{t=1}^n|c_{k,t}|^2 \leq nP$ with $P$ being the per-symbol maximum transmit power. $z_{t,m}$ is the additive white Gaussian noise (AWGN) with zero mean and $\sigma^2$ variance. Assume that the noise power $N_0=2\sigma^2$. Subsequently, the energy-per-bit $E_b$ over $N_0$, energy efficiency in coded systems, can be derived as $E_b/N_0=nP/(2b\sigma^2)$.

Commonly, the performance of unsourced random access is measured by \textit{per-user probability of misdetection} $P_{\textrm{MD}}$ and \textit{probability of false alarm} $P_{\textrm{FA}}$, i.e., the probability that the message was sent but not detected and the message was not sent but detected, expressed, respectively, as
\begin{equation}
P_{\textrm{MD}}=\frac{1}{K_a}\sum\limits_{k \in \mathcal{K}_a}\mathbb{P}(\textbf{m}_k \notin \hat{\mathcal{L}})
\end{equation}
and
\begin{equation}
P_{\textrm{FA}}=\frac{|\hat{\mathcal{L}}\backslash{\textbf{m}_k:k \in \mathcal{K}_a}|_H}{|\hat{\mathcal{L}}|_H}
\end{equation}
with $|\cdot|_H$ denoting the Hamming weight and $\hat{\mathcal{L}}$ presenting the list of transmitted messages recovered by the BS. Notice that the above two probabilities only depend on the active device set $\mathcal{K}_a$ and are independent of the total number of UEs $K_{\textrm{tot}}$.

\section{Design of Massive Unsourced Random Access Scheme}
\begin{figure}[h] \centering
\includegraphics [width=0.48\textwidth] {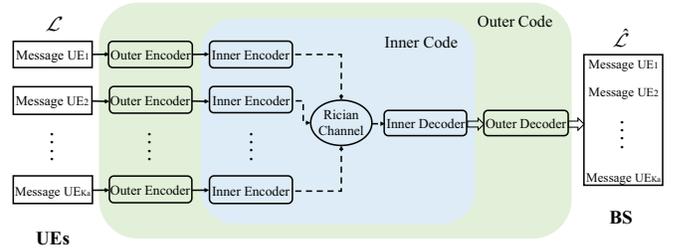}
\caption {The block diagram of unsourced random access.} \label{Fig1}
\end{figure}

For unsourced random access, a message will be transmitted in the uplink after being mapped to a codeword of a common codebook. Given the length of message $b$, the required number of codewords is not less than $2^b$. Even with short message for IoT applications, e.g., tens of bits, the number of codewords is prohibitively large, resulting in high computational complexity for codeword activity detection. To reduce the computational complexity, the $b$-bit message produced by an active UE will be divided into $L$ short fragments such that a small codebook is employed for all short fragments. Since these short fragments cannot represent complete information, they need to be stitched together based on some redundant check information for recovering the original messages sent by the active UEs. Therefore, for unsourced random access, the message transmission process usually consists of two-layer channel code, i.e., an outer code and an inner code, as shown in Fig. \ref{Fig1}. Specifically, the outer code aims to segment messages at the transmitter and stitch them together at the receiver via outer encoding and outer decoding, respectively. The inner code completes the codeword mapping at the transmitter and recovers the list of transmitted codewords at the receiver through inner encoding and inner decoding, respectively. In what follows, we design a comprehensive massive unsourced random access scheme for mMTC over Rician fading channels, including outer encoding, inner encoding, inner decoding, and outer decoding.

\subsection{Outer Encoding}
\begin{figure}[h] \centering
\includegraphics [width=0.48\textwidth] {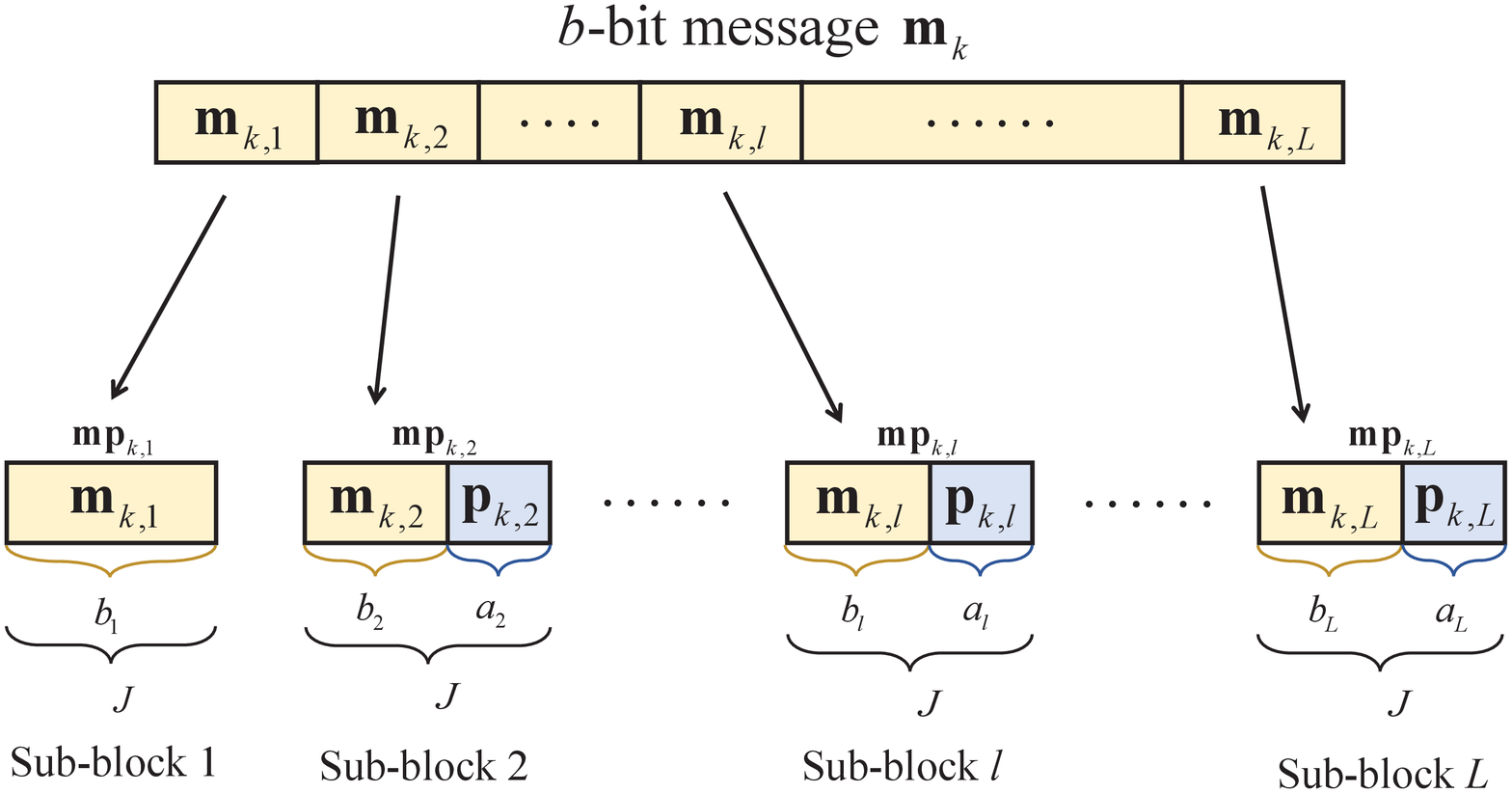}
\caption {The illustration of the proposed outer encoding scheme.} \label{Fig2}
\end{figure}

By outer encoding, the $b$-bit message $\{\textbf{m}_k, k \in \mathcal{K}_a\}$ of the $k$-th active UE is divided into $L$ fragments, i.e., $\{\textbf{m}_{k,1},\textbf{m}_{k,2},...,\textbf{m}_{k,L}\}$, and the length of the $l$-th fragment $\textbf{m}_{k,l}$ is $b_l$ satisfying $\sum_l b_l=b$. To facilitate the subsequent stitching in the outer decoding, the outer encoder needs to add several redundant parity check bits to the end of each fragment to form a fixed-length sub-block. Let define $\textbf{p}_{k,l}$ as the parity bits of sub-block $l$ with the length of $a_l$, which is obtained by combining the message bits of previous sub-block linearly, i.e.,
\begin{equation}
\textbf{p}_{k,l}^T=\left\{
\begin{array}{ll}
\textmd{0}, & l=1,\\
\sum\limits_{i=1}^{l-1}\textbf{m}_{k,i}^T\mathcal{G}_{i,l-1},\ \forall k, & l = 2,...,L.
\end{array}
\right.
\end{equation}
where $\mathcal{G}_{i,l-1}$ is the check bits generation matrix of size $b_i \times a_l$ and its elements follow Bernoulli distribution independently. Accordingly, the total length of a sub-block $\textbf{mp}_{k,l}$ is $J=b_l+a_l$ bits. The illustration of the outer encoding scheme is shown in Fig. \ref{Fig2}.

\subsection{Inner Encoding}
After the outer encoding, the collection of sub-blocks of all active UEs will be transmitted over multiple sub-slots in sequence. Specifically, the inner encoder maps each sub-block to a codeword of a common codebook. On the other hand, a transmission slot is partitioned into $L$ sub-slots of length $n_0=n/L$ and each active UE sends the $L$ codewords matched by $L$ sub-blocks from the common codebook across the $L$ sub-slots. Let $\textbf{C}=[\textbf{c}_1,...,\textbf{c}_{2^J}]\in \mathbb{C}^{n_0\times2^J}$ be the codebook with each column $\{\textbf{c}_i\in \mathbb{C}^{n_0\times1},\ i\in [1:2^J]\}$ representing a codeword. For a $J$-bit binary information sub-block, there are $2^J$ possible combinations, which are equal to the number of columns of the codebook. Then, the received signal $\textbf{Y}_l \in \mathbb{C}^{n_0 \times M}$ at the BS in sub-slot $l$ can be expressed as
\begin{eqnarray}
\textbf{Y}_l\!\!\!\!&=&\!\!\!\!\sum\limits_{k\in \mathcal{K}_a}\sqrt{\beta_k}\textbf{c}_{k,l}\textbf{h}^T_{k}+\textbf{Z}_l\\\nonumber
&=&\!\!\!\!\textbf{C}\textbf{A}_l\textbf{B}^{1/2}\textbf{H}+\textbf{Z}_l,\ \forall l\in\{1,...,L\},
\end{eqnarray}
where $\textbf{c}_{k,l}$ denotes the codeword mapped by the transmitted sub-block of the $k$-th active UE in the $l$-th sub-slot, $\textbf{B}=\textmd{diag}(\beta_1,...,\beta_{K_{\textrm{tot}}})$ is the large-scale fading coefficient, $\textbf{H}=[\textbf{h}_1,\textbf{h}_2,...,\textbf{h}_{K_{\textrm{tot}}}]^T \in \mathbb{C}^{K_{\textrm{tot}} \times M}$ is the channel matrix, and $\textbf{Z}_l$ represents the AWGN. $\textbf{A}_l \in \{0,1\}^{2^J \times K_{\textrm{tot}}}$ is the binary codeword activity indicator matrix. Its $k$-th column is given by $\textbf{a}_{l,k}=\textbf{e}_k$, which is a standard basis vector of dimension $2^J$ with zeros everywhere but a single one at ($\textmd{decimal}(\textbf{mp}_{k,l})+1$)-th element, if the $k$-th UE is active. Otherwise, $\textbf{a}_{l,k}=\textbf{0}$.

By letting $\gamma_{l,r}=\sum_k\beta_k a_{r,k}$, where $a_{r,k}$ is the $(r,k)$-th element of $\textbf{A}_l$, i.e., weighted summing each row of codeword activity indicator matrix $\textbf{A}_l$ with the large-scale fading coefficient as weight, the received signal at the BS can be rewritten more compactly as
\begin{equation}
\textbf{Y}_l=\textbf{C}\boldsymbol{\Gamma}_l^{1/2}\tilde{\textbf{H}}+\textbf{Z}_l,\ \forall l\in\{1,...,L\},
\end{equation}
where $\tilde{\textbf{H}}=[\tilde{\textbf{h}}_1,\tilde{\textbf{h}}_2,...,\tilde{\textbf{h}}_{2^J}]^T \in \mathbb{C}^{2^J \times M}$ is the channel matrix with transformed dimensions, and $\tilde{\textbf{h}}_r=\sum\limits_{k\in \mathcal{K}_r}\textbf{h}_k$ with $r \in [1:2^J], k\in [1:K_{\textrm{tot}}]$, which means that the channel related to the $r$-th codeword is the superposition of the channels of a set of UEs, denoted by $\mathcal{K}_r$, who simultaneously send the $r$-th codeword in the codebook. In this context, $\boldsymbol{\Gamma}_l=\textmd{diag}(\gamma_{l,1},...\gamma_{l,2^J})$ can be called the non-binary codeword activity matrix. Notice that the probability that $\gamma_{l,r}$ is non-zero is $1-(1-1/2^J)^{K_a}$, therefore vector $\boldsymbol{\gamma}_l$ is sparse.

\subsection{Inner Decoding}

\subsubsection{The estimation of codeword activity vector}
Once receiving the sub-slot signals, the BS recovers the transmitted codewords by performing the inner decoding. Intuitively, the recovery of lists of transmitted codewords is equivalent to the recovery of codeword activity matrix $\boldsymbol{\Gamma}_l$, which is also equivalent to detecting the sparse vector $\boldsymbol{\gamma}_l$. In this part, the inner decoding is performed based on the received signal $\textbf{Y}_l$ by exploiting the ML estimation method. Since the detection process is identical in each sub-slot, we drop the sub-slot index for notational simplicity.
In a certain sub-slot, with the deterministic $\textbf{C}$ and $\boldsymbol{\Gamma}$, the column vectors $\textbf{y}_m \in \mathbb{C}^{n_0 \times 1},\ m\in\{1,...,M\}$, of $\textbf{Y}$ can be equivalently regarded as independent samples following complex Gaussian distribution, i.e., $\textbf{y}_m \sim \mathcal{CN}(\boldsymbol{\mu}_m,\boldsymbol{\Sigma})$. The mean matrix $\boldsymbol{\mu}_m$ and covariance matrix $\boldsymbol{\Sigma}$ can be computed by
\begin{equation}
\boldsymbol{\mu}_m=\mathbb{E}[\textbf{y}_m]=\mathbb{E}[\textbf{C}\boldsymbol{\Gamma}^{1/2}\tilde{\textbf{H}}_{:,m}]=\textbf{C}\boldsymbol{\Gamma}^{1/2}\mathbb{E}[\tilde{\textbf{H}}_{:,m}]
\end{equation}
and
\begin{eqnarray}
\boldsymbol{\Sigma}&=&\mathbb{E}[\textbf{y}_m\textbf{y}_m^H]\\\nonumber
&=&\mathbb{E}[\textbf{C}\boldsymbol{\Gamma}^{1/2}\tilde{\textbf{H}}_{:,m}\tilde{\textbf{H}}_{:,m}^H\boldsymbol{\Gamma}^{1/2}\textbf{C}^H+\textbf{z}_m\textbf{z}_m^H]\\\nonumber
&=&\textbf{C}\boldsymbol{\Gamma}^{1/2}\mathbb{E}[\tilde{\textbf{H}}_{:,m}\tilde{\textbf{H}}_{:,m}^H]\boldsymbol{\Gamma}^{1/2}\textbf{C}^H+\sigma^2\textbf{I}_{n_0},
\end{eqnarray}
respectively. Based on the mean vector and covariance matrix of $\textbf{h}_k$ expressed in (\ref{hDistri}), the distribution of $\tilde{\textbf{H}}_{:,m}$ is given by
\begin{equation}
\tilde{\textbf{H}}_{:,m} \sim \mathcal{CN}(\tilde{\textbf{g}}_m,\textbf{R})
\end{equation}
with $\tilde{\textbf{g}}_m\!\!=\!\!\tilde{\textbf{G}}_{:,m}$ and $\textbf{R}\!\!=\!\!\textmd{diag}(\alpha'_1,...,\alpha'_{2^J})$. $\tilde{\textbf{G}}\!\!=\!\![\textbf{g}'_1,..,\textbf{g}'_{2^J}]^T\in \mathbb{C}^{2^J \times M}$ is the mean matrix of $\tilde{\textbf{H}}$. Thus, $\boldsymbol{\mu}_m$ and $\boldsymbol{\Sigma}$ can be further simplified to
\begin{equation}
\boldsymbol{\mu}_m\!=\textbf{C}\boldsymbol{\Gamma}^{1/2}\tilde{\textbf{g}}_m
\end{equation}
and
\begin{equation}
\boldsymbol{\Sigma}=\textbf{C}\boldsymbol{\Gamma}^{1/2}\textbf{R}\boldsymbol{\Gamma}^{1/2}\textbf{C}^H+\sigma^2\textbf{I}_{n_0}.
\end{equation}

Consequently, the likelihood function, which is the conditional probability of $\textbf{Y}$ given $\boldsymbol{\gamma}$, can be expressed as
\begin{eqnarray}
\mathbb{P}(\textbf{Y}|\boldsymbol{\gamma})&=&\mathbb{P}(\textbf{y}_1,...,\textbf{y}_M|\boldsymbol{\gamma})=\prod\limits_{m=1}^M \mathbb{P}(\textbf{y}_m|\boldsymbol{\gamma})\\\nonumber
&=&\prod\limits_{m=1}^M \frac{1}{|\pi\boldsymbol{\Sigma}|} \exp[-(\textbf{y}_m-\boldsymbol{\mu}_m)^H\boldsymbol{\Sigma}^{-1}(\textbf{y}_m-\boldsymbol{\mu}_m)].
\end{eqnarray}
Due to
\begin{eqnarray}
(\textbf{y}_m-\boldsymbol{\mu}_m)^H\boldsymbol{\Sigma}^{-1}(\textbf{y}_m-\boldsymbol{\mu}_m)\!\!\!\!&=&\!\!\!\!\textbf{y}_m^H\boldsymbol{\Sigma}^{-1}\textbf{y}_m-2\textbf{y}_m^H\boldsymbol{\Sigma}^{-1}\boldsymbol{\mu}_m\nonumber\\&+&\!\!\!\!\boldsymbol{\mu}_m^H\boldsymbol{\Sigma}^{-1}\boldsymbol{\mu}_m
\end{eqnarray}
and
\begin{equation}
\sum\limits_{m=1}^M \textbf{y}_m^H\boldsymbol{\Sigma}^{-1}\textbf{y}_m=\textmd{tr}(\boldsymbol{\Sigma}^{-1}\textbf{Y}\textbf{Y}^H),
\end{equation}
the likelihood function can be rewritten as
\begin{eqnarray}
\mathbb{P}(\textbf{Y}|\boldsymbol{\gamma})\!\!\!\!&=&\!\!\!\!\frac{1}{|\pi\boldsymbol{\Sigma}|^M} \exp\{-[\textmd{tr}(\boldsymbol{\Sigma}^{-1}\textbf{Y}\textbf{Y}^H)+\textmd{tr}(\boldsymbol{\Sigma}^{-1}\textbf{E}\textbf{E}^H)\nonumber\\&-&\!\!\!\!2\textmd{tr}(\boldsymbol{\Sigma}^{-1}\textbf{E}\textbf{Y}^H)]\},
\end{eqnarray}
where $\textbf{E}=[\boldsymbol{\mu}_1,...,\boldsymbol{\mu}_M] \in \mathbb{C}^{n_0 \times M}$. For convenience, we take the logarithm of the likelihood function. Then, maximizing the log-likelihood cost function $\textmd{log}(\mathbb{P}(\textbf{Y}|\boldsymbol{\gamma}))$ can be transformed into minimizing $-(1/M)\textmd{log}[\mathbb{P}(\textbf{Y}|\boldsymbol{\gamma})]$, which can be expressed as
\begin{equation}\label{OPinit}
\mathop{\textmd{min}}\limits_{\boldsymbol{\gamma} \in \mathbb{R}_+^{2^J}}\ 
\textmd{log}|\boldsymbol{\Sigma}|+\textmd{tr}(\boldsymbol{\Sigma}^{-1}\hat{\boldsymbol{\Sigma}})+\frac{1}{M}\textmd{tr}(\boldsymbol{\Sigma}^{-1}\textbf{E}\textbf{E}^H)-\frac{2}{M}\textmd{tr}(\boldsymbol{\Sigma}^{-1}\textbf{E}\textbf{Y}^H),
\end{equation}
where the empirical covariance matrix $\hat{\boldsymbol{\Sigma}}$ of column vectors of $\textbf{Y}$ is defined as
\begin{equation}
\hat{\boldsymbol{\Sigma}}=\frac{1}{M}(\textbf{Y}\textbf{Y}^H)=\frac{1}{M}\sum\limits_{m=1}^M\textbf{y}_m\textbf{y}_m^H.
\end{equation}

In contrast to Rayleigh fading channels, the coupling of variables due to the observations with non-zero mean causes the above problem (\ref{OPinit}) to be non-convex, which hinders us to obtain a solution directly. Fortunately, it is proved in \cite{AD1} that $\textmd{log}|\boldsymbol{\Sigma}|+\textmd{tr}(\boldsymbol{\Sigma}^{-1}\hat{\boldsymbol{\Sigma}})$ has a global minimizer which can be found by using coordinate descent approach (Theorem 2 in \cite{AD1}), which is called geodesic convexity. Inspired by this, we define a geodesic-convex function \cite{UN3} as
\begin{equation}
p(\boldsymbol{\gamma})=\textmd{log}|\boldsymbol{\Sigma}|+\textmd{tr}(\boldsymbol{\Sigma}^{-1}\hat{\boldsymbol{\Sigma}})
\end{equation}
and a non-convex function as
\begin{equation}
q(\boldsymbol{\gamma})=\frac{1}{M}\textmd{tr}(\boldsymbol{\Sigma}^{-1}\textbf{E}\textbf{E}^H)-\frac{2}{M}\textmd{tr}(\boldsymbol{\Sigma}^{-1}\textbf{E}\textbf{Y}^H).
\end{equation}
Since the multiplicative transformation of a sparse vector will not affect its sparsity and so will not affect its recovery, by letting $\tilde{\boldsymbol{\Gamma}}=\boldsymbol{\Gamma}^{1/2}\textbf{R}\boldsymbol{\Gamma}^{1/2}$ and replacing the variable $\boldsymbol{\gamma}$ with $\tilde{\boldsymbol{\gamma}}$, the above two functions are firstly reasonably approximated as
\begin{equation}
p(\tilde{\boldsymbol{\gamma}})=\textmd{log}|\tilde{\boldsymbol{\Sigma}}|+\textmd{tr}(\tilde{\boldsymbol{\Sigma}}^{-1}\hat{\boldsymbol{\Sigma}})
\end{equation}
and
\begin{equation}
q(\tilde{\boldsymbol{\gamma}})=\frac{1}{M}\textmd{tr}[\tilde{\boldsymbol{\Sigma}}^{-1}\textbf{C}\tilde{\boldsymbol{\Gamma}} \tilde{\textbf{G}}(\textbf{C}\tilde{\boldsymbol{\Gamma}} \tilde{\textbf{G}})^H]-\frac{2}{M}\textmd{tr}(\tilde{\boldsymbol{\Sigma}}^{-1}\textbf{C}\tilde{\boldsymbol{\Gamma}} \tilde{\textbf{G}}\textbf{Y}^H),
\end{equation}
where $\tilde{\boldsymbol{\Sigma}}=\textbf{C}\tilde{\boldsymbol{\Gamma}}\textbf{C}^H+\sigma^2\textbf{I}_{n_0}$ and $\tilde{\boldsymbol{\Gamma}}=\textmd{diag}(\tilde{\boldsymbol{\gamma}})$. Then the original problem (\ref{OPinit}) can be equivalently expressed as
\begin{equation}\label{log-likeli}
\mathop{\textmd{min}}\limits_{\tilde{\boldsymbol{\gamma}} \in \mathbb{R}_+^{2^J}}\ -\frac{1}{M}\textmd{log}[\mathbb{P}(\textbf{Y}|\tilde{\boldsymbol{\gamma}})]=p(\tilde{\boldsymbol{\gamma}})+q(\tilde{\boldsymbol{\gamma}}).
\end{equation}
To address problem (\ref{log-likeli}), we leverage a technique called Concave-Convex Procedure (CCCP) to obtain a suboptimal solution \cite{CCCP1}. Specifically, we construct an affine approximation of the non-convex function $q(\tilde{\boldsymbol{\gamma}})$, i.e., $q(\tilde{\boldsymbol{\gamma}})$ is linearized around a linearization point in each iteration. In the first iteration, any arbitrary $\tilde{\boldsymbol{\gamma}}^i$ in the feasible domain of the problem is selected and $q(\tilde{\boldsymbol{\gamma}})$ is approximated by an affine function through Taylor expansion around the point $\tilde{\boldsymbol{\gamma}}^i$. The first-order Taylor expansion of $q(\tilde{\boldsymbol{\gamma}})$ is given by
\begin{equation}\label{appr}
q(\tilde{\boldsymbol{\gamma}}) \approx q(\tilde{\boldsymbol{\gamma}}^i)+\triangledown q(\tilde{\boldsymbol{\gamma}}^i)^T(\tilde{\boldsymbol{\gamma}}-\tilde{\boldsymbol{\gamma}}^i).
\end{equation}
Replacing $q(\tilde{\boldsymbol{\gamma}})$ in problem (\ref{log-likeli}) with (\ref{appr}), we can derive the following solvable optimization problem
\begin{equation}
\mathop{\textmd{min}}\limits_{\tilde{\boldsymbol{\gamma}} \in \mathbb{R}_+^{2^J}}\ o(\tilde{\boldsymbol{\gamma}})=p(\tilde{\boldsymbol{\gamma}})+q(\tilde{\boldsymbol{\gamma}}^i)+\triangledown q(\tilde{\boldsymbol{\gamma}}^i)^T(\tilde{\boldsymbol{\gamma}}-\tilde{\boldsymbol{\gamma}}^i).\label{OPfinal}
\end{equation}

Now, the differentiable convex problem (\ref{OPfinal}) can be efficiently addressed by employing the coordinate descent algorithm \cite{CD1}. Let define a scalar function $s_r(d)=o(\tilde{\boldsymbol{\gamma}}+d\textbf{e}_r)$ on variable $d$ for each coordinate $r \in [1:2^J]$. $o(\cdot)$ is the log-likelihood function defined in (\ref{OPfinal}) and $\textbf{e}_r$ is the standard basis vector with a single one at position $r$ and zeros everywhere. Substituting $\tilde{\boldsymbol{\gamma}}+d\textbf{e}_r$ into (\ref{OPfinal}), the scalar function can be written as
\begin{equation}\label{scalar}
s_r(d)=p(\tilde{\boldsymbol{\gamma}}+d\textbf{e}_r)+q(\tilde{\boldsymbol{\gamma}}^i)+\triangledown q(\tilde{\boldsymbol{\gamma}}^i)^T(\tilde{\boldsymbol{\gamma}}+d\textbf{e}_r-\tilde{\boldsymbol{\gamma}}^i).
\end{equation}
Given $\tilde{\boldsymbol{\Sigma}}(\tilde{\boldsymbol{\gamma}})=\textbf{C}\tilde{\boldsymbol{\Gamma}}\textbf{C}^H+\sigma^2\textbf{I}_{n_0}$ and $\tilde{\boldsymbol{\Gamma}}=\textmd{diag}(\tilde{\boldsymbol{\gamma}})$, the first term at the right hand side of (\ref{scalar}) is given by
\begin{eqnarray}
p(\tilde{\boldsymbol{\gamma}}+d\textbf{e}_r)\!\!\!\!&=&\!\!\!\!\!\textmd{log}|\tilde{\boldsymbol{\Sigma}}(\tilde{\boldsymbol{\gamma}}+d\textbf{e}_r)|+\textmd{tr}(\tilde{\boldsymbol{\Sigma}}^{-1}(\tilde{\boldsymbol{\gamma}}+d\textbf{e}_r)\hat{\boldsymbol{\Sigma}})\nonumber\\
&=&\!\!\!\!\!\textmd{log}|\tilde{\boldsymbol{\Sigma}}+d\textbf{c}_r\textbf{c}_r^H|+\textmd{tr}[(\tilde{\boldsymbol{\Sigma}}+d\textbf{c}_r\textbf{c}_r^H)^{-1}\hat{\boldsymbol{\Sigma}}].
\end{eqnarray}
According to the Sherman-Morrison rank-1 update property \cite{CD2}, we have
\begin{equation}
|\tilde{\boldsymbol{\Sigma}}+d\textbf{c}_r\textbf{c}_r^H|=(1+d\textbf{c}_r^H\tilde{\boldsymbol{\Sigma}}^{-1}\textbf{c}_r)|\tilde{\boldsymbol{\Sigma}}|
\end{equation}
and
\begin{equation}
(\tilde{\boldsymbol{\Sigma}}+d\textbf{c}_r\textbf{c}_r^H)^{-1}=\tilde{\boldsymbol{\Sigma}}^{-1}-\frac{d\tilde{\boldsymbol{\Sigma}}^{-1}\textbf{c}_r\textbf{c}_r^H\tilde{\boldsymbol{\Sigma}}^{-1}}{1+d\textbf{c}_r^H\tilde{\boldsymbol{\Sigma}}^{-1}\textbf{c}_r}.
\end{equation}
Therefore, $s_r(d)$ can be rewritten as
\begin{eqnarray}
s_r(d)\!\!\!\!&=&\!\!\!\!\textmd{log}(1+d\textbf{c}_r^H\tilde{\boldsymbol{\Sigma}}^{-1}\textbf{c}_r)-\frac{d\textbf{c}_r^H\tilde{\boldsymbol{\Sigma}}^{-1}\hat{\boldsymbol{\Sigma}}\tilde{\boldsymbol{\Sigma}}^{-1}\textbf{c}_r}{1+d\textbf{c}_r^H\tilde{\boldsymbol{\Sigma}}^{-1}\textbf{c}_r}\nonumber\\&&\!\!\!\!+\triangledown q(\tilde{\boldsymbol{\gamma}}^i)^T(d\textbf{e}_r)+\mathfrak{c},
\end{eqnarray}
where the term $\mathfrak{c}=p(\tilde{\boldsymbol{\gamma}})+q(\tilde{\boldsymbol{\gamma}}^i)+\triangledown q(\tilde{\boldsymbol{\gamma}}^i)^T(\tilde{\boldsymbol{\gamma}}-\tilde{\boldsymbol{\gamma}}^i)$ is a constant independent of $d$.
Taking the first-order derivative of the function $s_r(d)$, we have
\begin{equation}
s_r'(d)=\frac{\textbf{c}_r^H\tilde{\boldsymbol{\Sigma}}^{-1}\textbf{c}_r}{1+d\textbf{c}_r^H\tilde{\boldsymbol{\Sigma}}^{-1}\textbf{c}_r}-\frac{\textbf{c}_r^H\tilde{\boldsymbol{\Sigma}}^{-1}\hat{\boldsymbol{\Sigma}}\tilde{\boldsymbol{\Sigma}}^{-1}\textbf{c}_r}{(1+d\textbf{c}_r^H\tilde{\boldsymbol{\Sigma}}^{-1}\textbf{c}_r)^2}+\triangledown q(\tilde{\boldsymbol{\gamma}}^i)^T\textbf{e}_r.
\end{equation}
By letting $s_r'(d)=0$, the solution can be derived as
\begin{equation}
d^*=\tfrac{(n_1^2-2n_1n_3)-\sqrt{(2n_1n_3-n_1^2)^2-4n_1^2n_3(n_2+n_3-n_1)}}{2n_1^2n_3},\label{dstar}
\end{equation}
where $n_1=\textbf{c}_r^H\tilde{\boldsymbol{\Sigma}}^{-1}\textbf{c}_r$, $n_2=\textbf{c}_r^H\tilde{\boldsymbol{\Sigma}}^{-1}\hat{\boldsymbol{\Sigma}}\tilde{\boldsymbol{\Sigma}}^{-1}\textbf{c}_r$, $n_3=-[\triangledown q(\tilde{\boldsymbol{\gamma}}^i)]_r$, and the derivation of $\triangledown q(\tilde{\boldsymbol{\gamma}})$ is shown in Appendix A. Note that the another root of the quadratic equation falls outside the domain of definition $d \in (-\frac{1}{\textbf{c}_r^H\tilde{\boldsymbol{\Sigma}}^{-1}\textbf{c}_r},\infty)$ and thus is discarded. Moreover, it is easy to find that $s_r(d)$ tends to infinity when $d$ approaches the two boundaries of the domain. Therefore, $d^*$ is indeed the minimum of $s_r(d)$. Finally, update $\gamma_r\leftarrow\gamma_r+d^*$ to obtain the optimal solution of problem (\ref{OPfinal}).

Next, we denote the optimal solution of problem (\ref{OPfinal}) in the first iteration as $\tilde{\boldsymbol{\gamma}}^{i+1}$. In each subsequent iteration, the problem (\ref{OPfinal}) is solved through a similar procedure, in which the Taylor expansion point is replaced by the optimal solution of last iteration. This iteration process continues until the objective function converges. Since the convergence behavior of the CCCP has been well proven in previous literatures \cite{CCCP1,CCCP2}, we only illustrate the convergence illustration directly in simulation section due to space limitation. In summary, the codeword activity detection algorithm can be described as Algorithm 1.

\begin{algorithm}
\caption{: Codeword Activity Detection}
\label{alg1}
\hspace*{0.02in} {\bf Input:}
 The sample covariance matrix $\hat{\boldsymbol{\Sigma}}=\frac{1}{M}\textbf{YY}^H$.\\
\hspace*{0.02in} {\bf Output:}
 The estimated $\tilde{\boldsymbol{\gamma}}$
\begin{algorithmic}[1]
\STATE{\textbf{Initialize} $\tilde{\boldsymbol{\Sigma}}=\sigma^2\textbf{I}_{n_0}$, iteration index $i=0$, $\tilde{\boldsymbol{\gamma}}^i=\textbf{0}$, convergence accuracy $\Delta=0.1$, the maximum number of iterations $I_{\max}=20$, the temporary variable $v=0$, and select the maximum number of traverse coordinates $T_{\max}>2^J$.}
\WHILE{$\Delta>0.01$ and $i\leq I_{\max}$}
 \STATE{Compute $v=o(\tilde{\boldsymbol{\gamma}}^i)$ according to (\ref{OPfinal});}
 \FOR{$j=1:T_{\max}$}
  \STATE{Select an index $r \in [1:2^J]$ randomly, which corresponds to the $r$-th element of $\tilde{\boldsymbol{\gamma}}^i$;}
  \STATE{Compute $d^*$ according to (\ref{dstar}); }
  \STATE{Set $d^*=\max\{d^*,-\tilde{\boldsymbol{\gamma}}^i_r\}$;}
  \STATE{Update $\tilde{\boldsymbol{\gamma}}^i_r \leftarrow \tilde{\boldsymbol{\gamma}}^i_r+d^*$;}
  \STATE{Update $\tilde{\boldsymbol{\Sigma}} \leftarrow \tilde{\boldsymbol{\Sigma}}+d^*(\textbf{c}_r\textbf{c}_r^H)$;}
 \ENDFOR
 \STATE{Obtain $\tilde{\boldsymbol{\gamma}}^{i+1}=\tilde{\boldsymbol{\gamma}}^i$, and compute $o(\tilde{\boldsymbol{\gamma}}^{i+1})$ according to (\ref{OPfinal});}
 \STATE{Update $\Delta=v-o(\tilde{\boldsymbol{\gamma}}^{i+1})$;}
 \STATE{Update $i=i+1$;}
\ENDWHILE
\end{algorithmic}
\end{algorithm}

\subsubsection{The recovery of transmitted sub-blocks}

Once the estimated codeword activity vector $\tilde{\boldsymbol{\gamma}}$, i.e., $\hat{\tilde{\boldsymbol{\gamma}}}$, is obtained, we set a small positive fluctuation $\delta$ to choose the sparsity parameter $K=\tilde{K}_a+\delta$. Note that the probabilities of misdetection and false alarm can be swung by setting different values of $\delta$. It is also worth pointing out that the BS does not know the exact number of active UEs in a cell and can only determine an approximate number based on the information provided by the data center such as the network operator, i.e., $\tilde{K}_a=\varepsilon K_t$, where $\varepsilon$ is the activity probability with $0\leq\varepsilon\leq1$. Apparently, the value of $\varepsilon$ in a cell with a high traffic load is usually greater than that in a isolated cell. Besides, the value of $\varepsilon$ during the day is always greater than that at night in the same service area. Then let the $K$ largest elements of $\hat{\tilde{\boldsymbol{\gamma}}}$ be one and the remaining $2^J-K$ elements be zero. Naturally, the $r$-th codeword in codebook $\textbf{C}$, where $\hat{\tilde{\boldsymbol{\gamma}}}_r$ is equal to one, is the transmitted codeword from one of active UEs, and the sub-block mapped back by this codeword is one of the outputs of the inner decoder. Note that although $K$ may be greater than the number of transmitted codewords, the extra codewords cannot satisfy all the parity checks in outer decoding. Thus, the impact of overestimation of the active codewords can be reduced.

\subsection{Outer Decoding}
\begin{figure}[h] \centering
\includegraphics [width=0.5\textwidth] {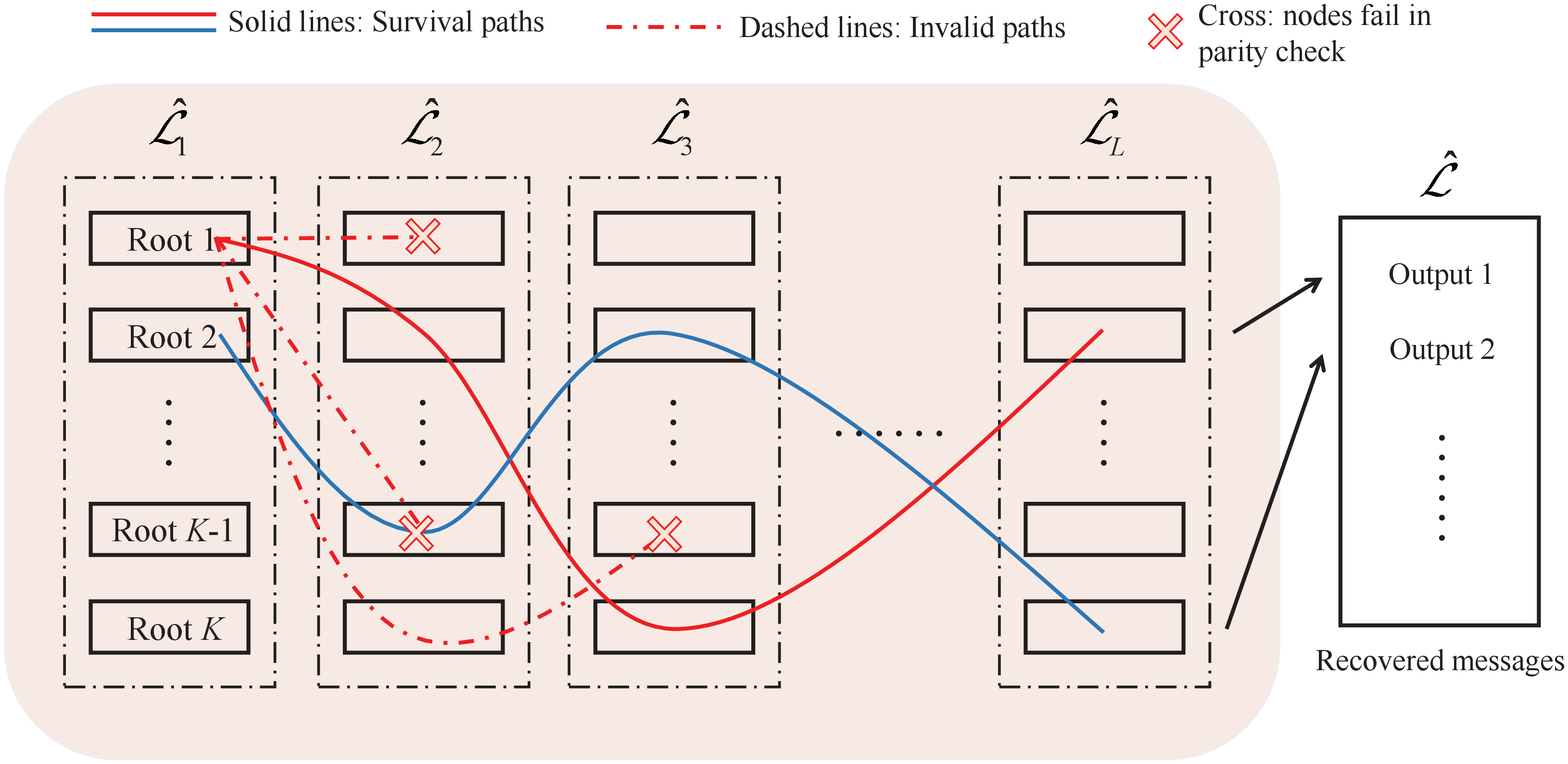}
\caption {A diagram of the proposed outer decoding scheme.} \label{Fig3}
\end{figure}

To recover the original messages sent from all active UEs, the outer decoder needs to stitch the sub-blocks bits together in $L$ sub-slots output by the inner decoder. Define the output lists of the inner decoder in the $l$-th sub-slot as $\hat{\mathcal{L}}_l, l\in\{1,...,L\}$, all sub-blocks in these lists are candidate messages, as shown in Fig. \ref{Fig3}. First, every sub-block in list $\hat{\mathcal{L}}_1$ is selected as the root node one by one. Once the root node from $\hat{\mathcal{L}}_1$ is fixed, there are $K$ possible nodes from $\hat{\mathcal{L}}_2$ for choice. Then in list $\hat{\mathcal{L}}_3$, there are $K^2$ possible choices. In other words, counting $K$ root nodes in list $\hat{\mathcal{L}}_1$, when reaching the last node from list $\hat{\mathcal{L}}_L$, there are a total of $K^L$ candidate paths as possible transmitted messages. Yet, not all paths will be connected. The paths that do not meet the parity check constraints, which are called invalid paths, will be discarded directly. When the last list $\hat{\mathcal{L}}_L$ is reached, if there is one and only one path that satisfies all parity checks, which is called survival path, this path is one of the output messages of the outer decoder after removing the check bits; Otherwise, outer decoding fails at this time. After that, repeat this process with the next root node until $K$ recovered messages are obtained. Note that the decoding complexity depends on the number of nodes that meet parity checks in each list during the traversal process, which has been quantified in \cite{UN2}.
Generally, since the number of check bits is constrained, the transmission delay introduced by redundancy is limited. Yet, the processing complexity can be decreased significantly through message segment, leading to the reduction of processing delay.

\section{Analysis and Optimization of Massive Unsourced Random Access}
In this section, to evaluate the performance of the proposed codeword activity detection algorithm for the inner code, a performance analysis from the perspective of estimation error is carried out. Meanwhile, for the outer code, the structure of sub-block, i.e., lengths of message bits and parity check bits, is optimized to strike a balance between the complexity and the error probability of the outer decoding.

\subsection{Performance Analysis of Inner Decoding}
To examine the effectiveness of the proposed codeword activity detection algorithm, we focus on the distribution of estimation error, $\hat{\tilde{\boldsymbol{\gamma}}}-\tilde{\boldsymbol{\gamma}}$, which can characterize the estimation accuracy of codeword activity vector $\tilde{\boldsymbol{\gamma}}$. Since it is difficult to analyze the estimation error from Algorithm 1 directly, as an approximate approach, we focus on the \emph{asymptotically unbiased property} of the maximum likelihood estimator (MLE), i.e., the estimated value $\hat{\tilde{\boldsymbol{\gamma}}}$ that maximizes the likelihood function. According to the classic estimation theory of statistical signal processing \cite{SP}, asymptotically unbiased property of MLE describes that if the probability density function (PDF) of received signals meets some certain conditions, the MLE $\hat{\tilde{\boldsymbol{\gamma}}}$ will tend to Gaussian distribution when there are enough received signals observations. For the proposed Algorithm 1, it is possible to obtain enough signals observations if the number of BS antennas $M$ is sufficiently large. In this context, the distribution of MLE can be expressed as
\begin{equation}\label{PDFMLE}
\hat{\tilde{\boldsymbol{\gamma}}} \sim^a \mathcal{N}(\tilde{\boldsymbol{\gamma}},\textbf{F}^{-1}(\tilde{\boldsymbol{\gamma}})), \ \ as\ M\rightarrow\infty
\end{equation}
where $\sim^a$ denotes the asymptotic distribution, $\textbf{F}(\tilde{\boldsymbol{\gamma}})$ is the Fisher Information Matrix whose $(i,j)$-th element is given by
\begin{equation}
[\textbf{F}(\tilde{\boldsymbol{\gamma}})]_{ij}=-\mathbb{E}[\frac{\partial^2\log\mathbb{P}(\textbf{Y}|\tilde{\boldsymbol{\gamma}})}{\partial\tilde{\boldsymbol{\gamma}}_i\partial\tilde{\boldsymbol{\gamma}}_j}],
\end{equation}
where $\log\mathbb{P}(\textbf{Y}|\tilde{\boldsymbol{\gamma}})$ is derived from (\ref{log-likeli}) and the expectation is taken over $\textbf{Y}$. Additionally, for (\ref{PDFMLE}) to hold true, the following conditions should be satisfied: (1) The first-order and the second-order derivatives of the log-likelihood function exist; (2) The Fisher Information Matrix is non-zero. To verify these conditions, we derive the Fisher Information Matrix in Appendix B indicating that these two conditions are automatically satisfied in Algorithm 1. Consequently, the asymptotic distribution of the estimation error $\hat{\tilde{\boldsymbol{\gamma}}}-\tilde{\boldsymbol{\gamma}}$ can be expressed as
\begin{equation}
\sqrt{n}(\hat{\tilde{\boldsymbol{\gamma}}}-\tilde{\boldsymbol{\gamma}}) \sim^a \mathcal{N}(\textbf{0},n\textbf{F}^{-1}(\tilde{\boldsymbol{\gamma}})), \ \ as\ M\rightarrow\infty
\end{equation}
where $n=M$ is the number of received signals observations. It is worth pointing out that the estimation error $\hat{\tilde{\boldsymbol{\gamma}}}-\tilde{\boldsymbol{\gamma}}$, which follows the multi-variate Gaussian distribution, is determined by not only its mean vector, but also the eigenvalues and eigenvectors of its covariance matrix. To be specific, the location of the error distribution depends on the mean vector and the rotation of it is controlled by the eigenvectors of the covariance matrix, while the scaling in each direction is controlled by the eigenvalues of the covariance matrix. Herein, the mean vector of the estimation error is $\textbf{0}$, so the center of the error distribution is at the origin of the coordinates. Later, the scatter plot of the error distribution and PDF diagram of the estimation error will be illustrated in simulation section to confirm the accuracy of the proposed asymptotic approximation.

\subsection{Length Optimization of Outer Encoding}
During the outer encoding, some parity check bits are introduced to message fragments to facilitate the stitching of sub-blocks in outer decoding. In general, the required computational time and error probability of the outer decoding are directly determined by the lengths of message bits and parity check bits in each sub-block. Given the packet length, the number of information bits and the number of check bits have a nontrivial tradeoff \cite{UN2}. On the one hand, the more parity check bits of the sub-block in the previous sub-slot, the shorter the required outer decoding time, since invalid paths can be found and discarded in the early stages, further leading to the reduction of the number of sub-blocks that need to be checked. However, this causes a higher probability of error. On the other hand, assigning more parity check bits to the sub-block in the later sub-slot will lower the probability of error but extend the required outer decoding time. Traditionally, fixed length allocation is adopted due to its simplicity. However, the fixed allocation scheme cannot balance error probability and decoding time. In the context of massive connectivity and Rician channel fading, it makes sense to design a length allocation scheme to minimize the decoding time, while ensuring the success of the outer decoding.

Prior to designing the length allocation scheme, we first discuss the situations that may cause the outer decoding to fail. As mentioned earlier, there are two types of failures, misdetection and false alarm. To be specific, misdetection means that the number of survival paths from a certain root node to the last sub-slot, denoted by $\chi_L$, is zero. Clearly, misdetection occurs since the correct sub-blocks that meet the check constraints are not included in the output lists of the inner decoder. Furthermore, there are two main circumstances that lead to the missing of correct sub-blocks. One is that the estimation of $\tilde{\boldsymbol{\gamma}}$ is inaccurate and the other is that the value of fluctuation $\delta$ is set too small. In contrast, false alarm means that the number of survival paths $\chi_L$ from a certain root node is greater than or equal to two. The reason for this event is that the value of fluctuation $\delta$ is set too large, resulting in too many irrelevant sub-blocks to be the output of the inner decoder such that a path containing wrong sub-blocks but satisfying all the checks to be survived.
Under the implementation and parameter setting of our detection algorithm, it is assumed reasonably that all the sub-blocks transmitted from active UEs appear on the output lists of the inner decoder, i.e., no correct sub-blocks are missing. In other words, the outer decoder fails only when there are two or more survival paths.
%
%
In this context, the length allocation of parity check bits can be designed to strike a balance between the complexity and the error probability of the outer decoding based on the results derived in \cite{UN2}. Subject to the probability constraints of decoding failure and length constraints, the optimization problem with decoding complexity $\Xi$ over Rician fading channels being the objective function can be formulated as OP1 at the top of this page,
\begin{figure*}
\begin{eqnarray}\label{OP1}
\textmd{OP1}:&\mathop{\min}\limits_{(a_2,...,a_L)\in\mathbb{N}_+^{L-1}}&\!\! \Xi=K(L-1)+K\sum\limits_{j=2}^{L-1}\sum\limits_{m=2}^jK^{j-m}(K-1)\prod_{l=m}^j(2^{-a_l})\\\nonumber
\textmd{s.t.} \!\!\!\!&\textmd{C1}:&\!\!\!\!0\leq a_l \leq J,\ l=2,...,L\\\nonumber
\!\!\!\!&\textmd{C2}:&\!\!\!\!\sum\limits_{l=2}^L a_l=L \times J -b,\\\nonumber
\!\!\!\!&\textmd{C3}:&\!\!\!\!P_{\textmd{outer}} \leq p_{\textmd{th}},
\end{eqnarray}
\hrulefill
\end{figure*}
where $P_{\textmd{outer}}$ and $p_{\textmd{th}}$ represent the probability of decoding failure in the presence of Rician channel fading and its tolerable threshold. C1, C2, and C3 are, respectively, constraints on the length of parity check bits in each sub-block, total length of parity check bits in all sub-blocks, and error probability.

Since the lengths of parity check bits are non-negative integers, it is challenging to tackle OP1 directly with tolerable complexity. As a result, the optimization variables $(a_2,...,a_L)\in \mathbb{N}_+^{L-1}$ are relaxed to real-valued continuous numbers, i.e., $(\tilde{a}_2,...,\tilde{a}_L)\in \mathbb{R}_+^{L-1}$. Meanwhile, $P_{\textmd{outer}}$ is replaced by its Markov upper bound $\mathbb{E}[\chi_L]$, which has a closed form expression \cite{UN2}. Then, by letting $2^{-\tilde{a}_l}=\tau_l$ and substituting variables $\tilde{a}_l=\log_2(1/\tau_l)$ into problem (\ref{OP1}), the original problem OP1 can be rewritten as OP1$'$ at the top of the next page, which is a geometric programming problem \cite{CVX}.
\begin{figure*}[ht]
\begin{eqnarray}
\textmd{OP1}':&\mathop{\min}\limits_{(\tau_2,...,\tau_L)\in \mathbb{R}_+^{L-1}}&\!\!K(L-1)+K\sum\limits_{j=2}^{L-1}\sum\limits_{m=2}^jK^{j-m}(K-1)\prod_{l=m}^j\tau_l\\\nonumber
\textmd{s.t.} \!\!\!\!&\textmd{C1}':&\!\!\!\!2^{-J}\leq \tau_l \leq 1,\ l=2,...,L,\\\nonumber
\!\!\!\!&\textmd{C2}':&\!\!\!\!\sum\limits_{l=2}^L \log_2(1/\tau_l)=L \times J -b,\\\nonumber
\!\!\!\!&\textmd{C3}':&\!\!\!\!\mathbb{E}[\chi_L] \leq p_{\textmd{th}}.
\end{eqnarray}
\hrulefill
\end{figure*}
Next, we transform OP1$'$ into a convex form by taking the logarithm. For the constraint $\textmd{C3}'$, we have
\begin{eqnarray}
&&\!\!\!\!\sum\limits_{m=2}^{L}\left(K^{L-m}(K-1)\prod_{l=m}^{L}\tau_l\right)=\sum\limits_{m=2}^{L}\left(c_m\prod_{l=m}^{L}\tau_l\right)\nonumber\\
&=&\!\!\!\!\sum\limits_{m=2}^{L}2^{\log_2(c_m)+\sum\limits_{l=m}^L\log_2(\tau_l)}\nonumber\\
&=&\!\!\!\!\sum\limits_{m=2}^{L}2^{\log_2(c_m)+\sum\limits_{l=m}^L\omega_l}
\end{eqnarray}
with $c_m=K^{L-m}(K-1)$ and $\omega_l=\log_2(\tau_l)=-a_l$. Similar transformations are also performed to the objective function, the final expression of which is given by
\begin{equation}
\Xi'=K(L-1)+K\sum\limits_{j=2}^{L-1}\sum\limits_{m=2}^{j}2^{\log_2(c'_{m,j})+\sum\limits_{l=m}^j\omega_l},
\end{equation}
where $c'_{m,j}=K^{j-m}(K-1)$. Hence, the optimization problem is a standard geometric program in convex form of variable $\omega_l$ and it can be handled efficiently by convex optimization solver.

Finally, with the optimal solution $\omega_l^*$, the optimized lengths of parity check bits can be given by $a_l=[-\omega_l^*]^N,\ l\in[2,L]$, where $[\cdot]^N$ means to take the nearest integer. Hence, we can allocate the length according to the requirements on performance and complexity of mMTC over Rician fading channels in 6G wireless networks.

\section{Numerical Results}
In this section, we conduct extensive simulations to validate the effectiveness of the proposed massive unsourced random access scheme in a single-cell 6G mMTC system. We assume that only about $10\%$ UEs are active at the same time due to the sporadic traffic. Unless further specified, the main simulation parameters are set as: $K_{\textmd{tot}}=1000$, $K_a=100$, $M=256$, $\sigma^2=1$, $b=54$ bits, $J=6$ bits, $L=20$, $\delta=10$, $n_0=100$, and $E_b/N_0=0\ \textmd{dB}$. In addition, the large-scale fading coefficient and Rician factor can be calculated as \cite{Rice3}
\begin{equation}
\beta_k=126+35\log_{10}(d_k),\ \ \kappa_k=10^{1.3-0.003d_k},
\end{equation}
where the distance between the $k$-th UE to the BS is $d_k \in (0,d_{\textrm{max}})\ \textmd{m}$ with $d_{\textrm{max}}=300$.

\begin{figure}[h] \centering
\includegraphics [width=0.5\textwidth] {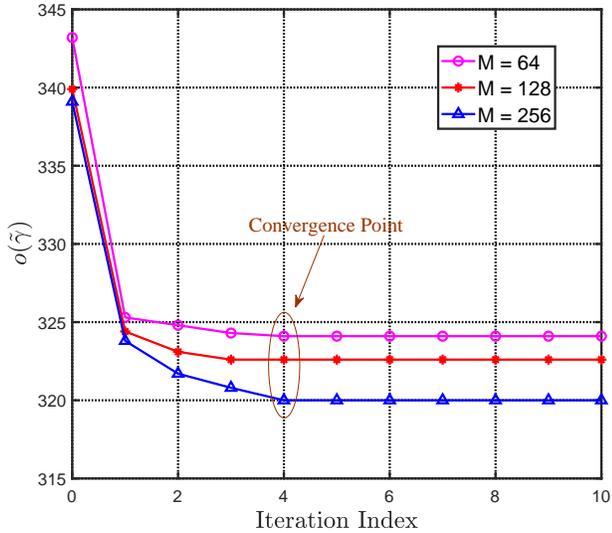}
\caption {The convergence speed of Algorithm 1 with different numbers of BS antennas. $K_a=100$.} \label{Fig4}
\end{figure}
We first verify the convergence behavior and computational complexity of Algorithm 1 with different numbers of BS antennas. As seen in Fig. \ref{Fig4}, the proposed codeword activity detection algorithm converges and the value of ML cost function $o(\tilde{\boldsymbol{\gamma}})$ decreases monotonically to its minimum value within 5 iterations in all the three cases, which demonstrates that Algorithm 1 has a fast convergence behavior. Additionally, since the ML optimization problem is converted to a convex one, it can be solved with affordable complexity related to the number of variables and constraints \cite{CVX}, which are both $2^J$ here, and independent of the number of antennas $M$. Thus, as long as the length of sub-block is not too large, the complexity of each iteration is tolerable for implementation. These facts all indicate the feasibility and efficiency of Algorithm 1.

\begin{figure}
\centering
\subfigure[ ]{\label{fig:subfig:a}\includegraphics [width=3.1in] {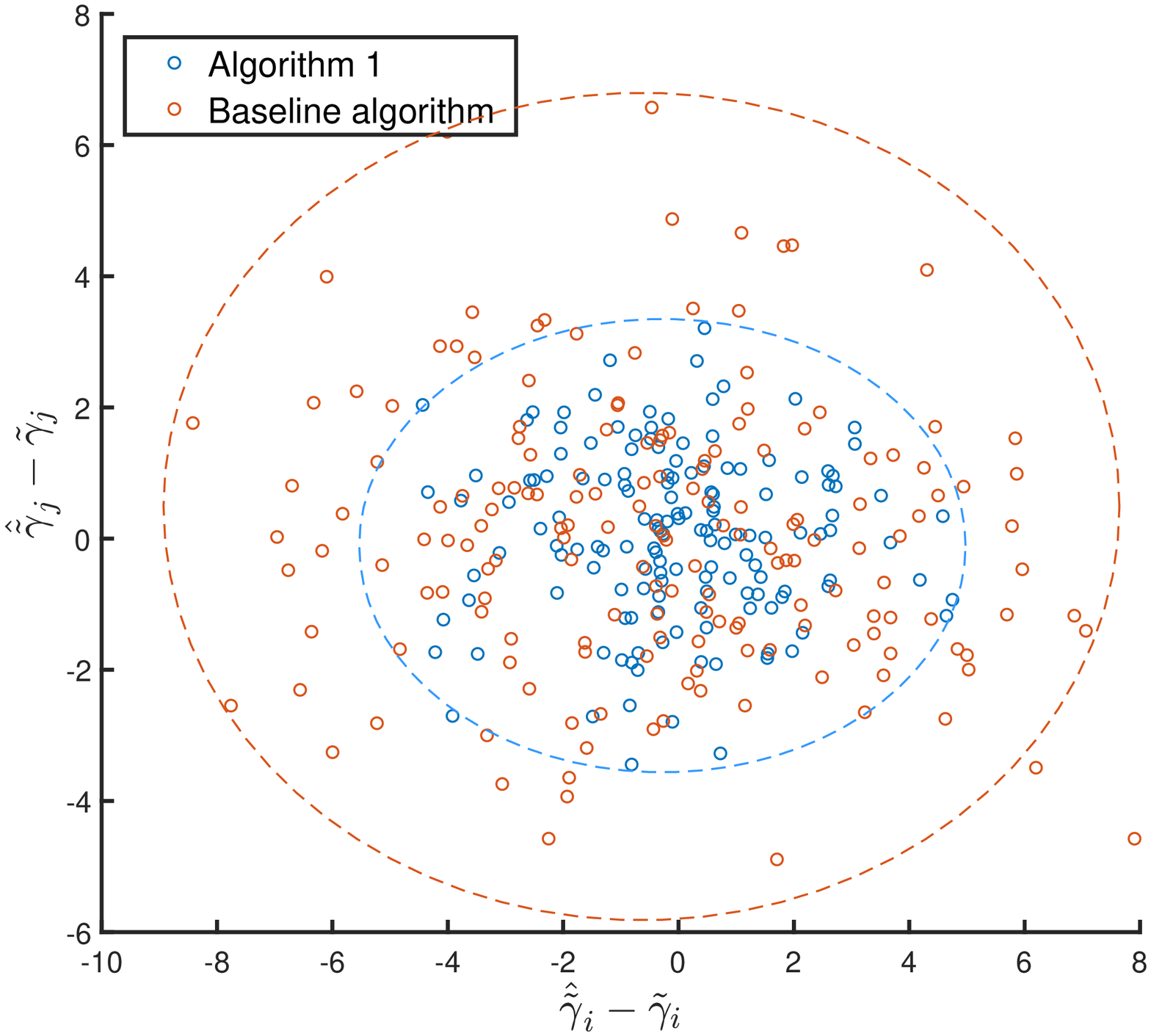}}
\subfigure[ ]{\label{fig:subfig:b}\includegraphics [width=3.3in] {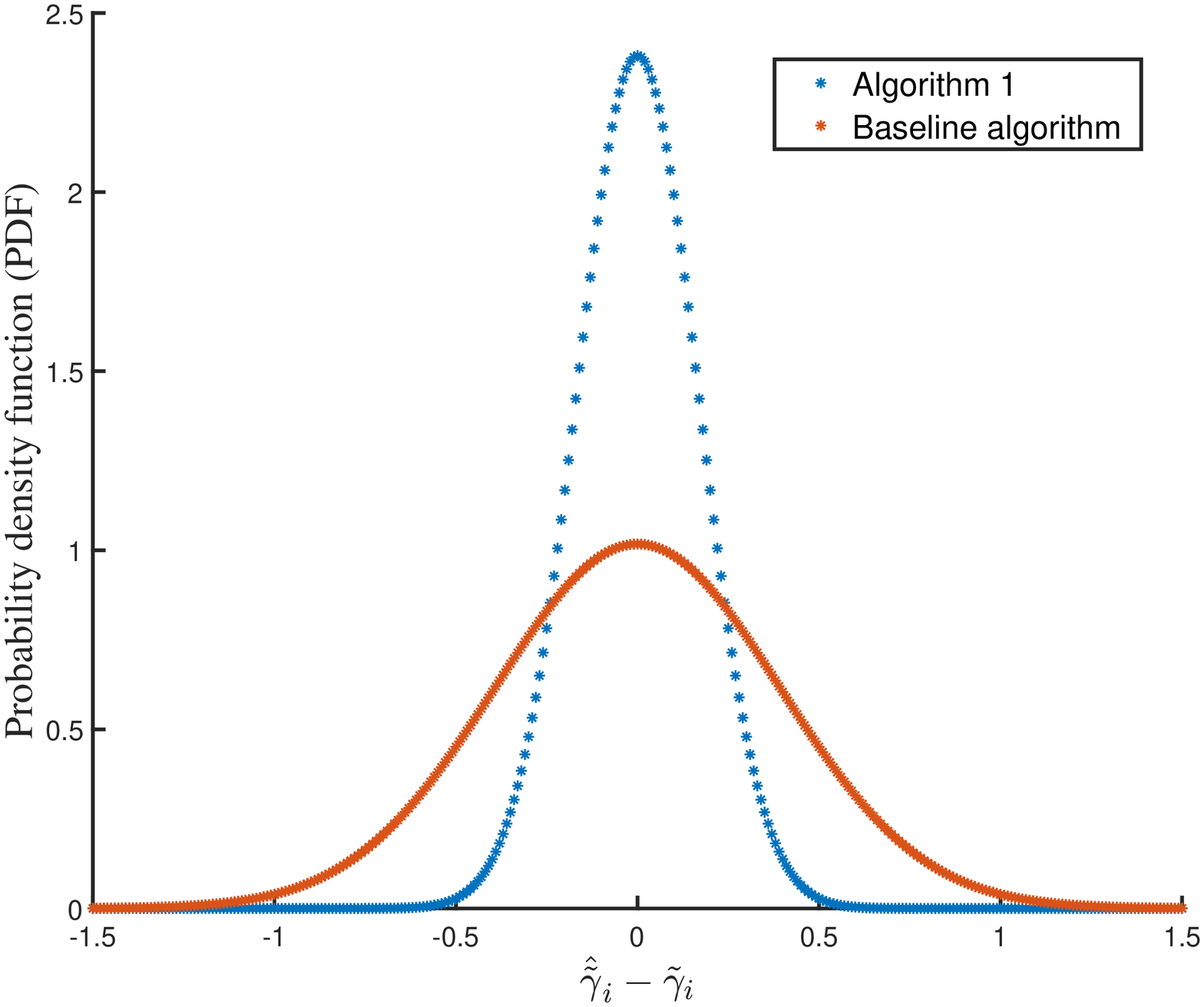}}
\caption{(a). The scatter plots of the error distribution; (b). the PDF diagrams of the estimation error for the proposed Algorithm 1 and the baseline algorithm. $K_a=100$, $M=128$.}
\label{Fig56:sub}
\end{figure}
Secondly, we show the performance difference between the proposed Algorithm 1 which exploits non-zero mean based ML detection and a baseline algorithm, i.e., the only covariance based ML detection algorithm which does not consider the LOS propagation component \cite{UN3}. Fig. \ref{Fig56:sub}. (a) and Fig. \ref{Fig56:sub}. (b) demonstrate the approximated distribution of estimation error of the codeword activity vector $\tilde{\boldsymbol{\gamma}}$ from two different perspectives. For simplicity, each coordinate of estimation error $\hat{\tilde{\boldsymbol{\gamma}}}-\tilde{\boldsymbol{\gamma}}$ is treated as independent and thus we plot the empirical distribution of the coordinate-wise error.
In Fig. \ref{Fig56:sub}. (a), it is seen that the scattered points of the estimation error of the proposed non-zero mean based detection algorithm are more concentrated. 
Similarly, in Fig. \ref{Fig56:sub}. (b), the PDF curve of the proposed Algorithm 1 is taller and thinner, which means the probability that the estimation error is distributed near the origin of the coordinates is higher. In a nutshell, Algorithm 1 is indeed capable of improving the estimation accuracy of codeword activity vector.

\begin{figure}[h] \centering
\includegraphics [width=0.5\textwidth] {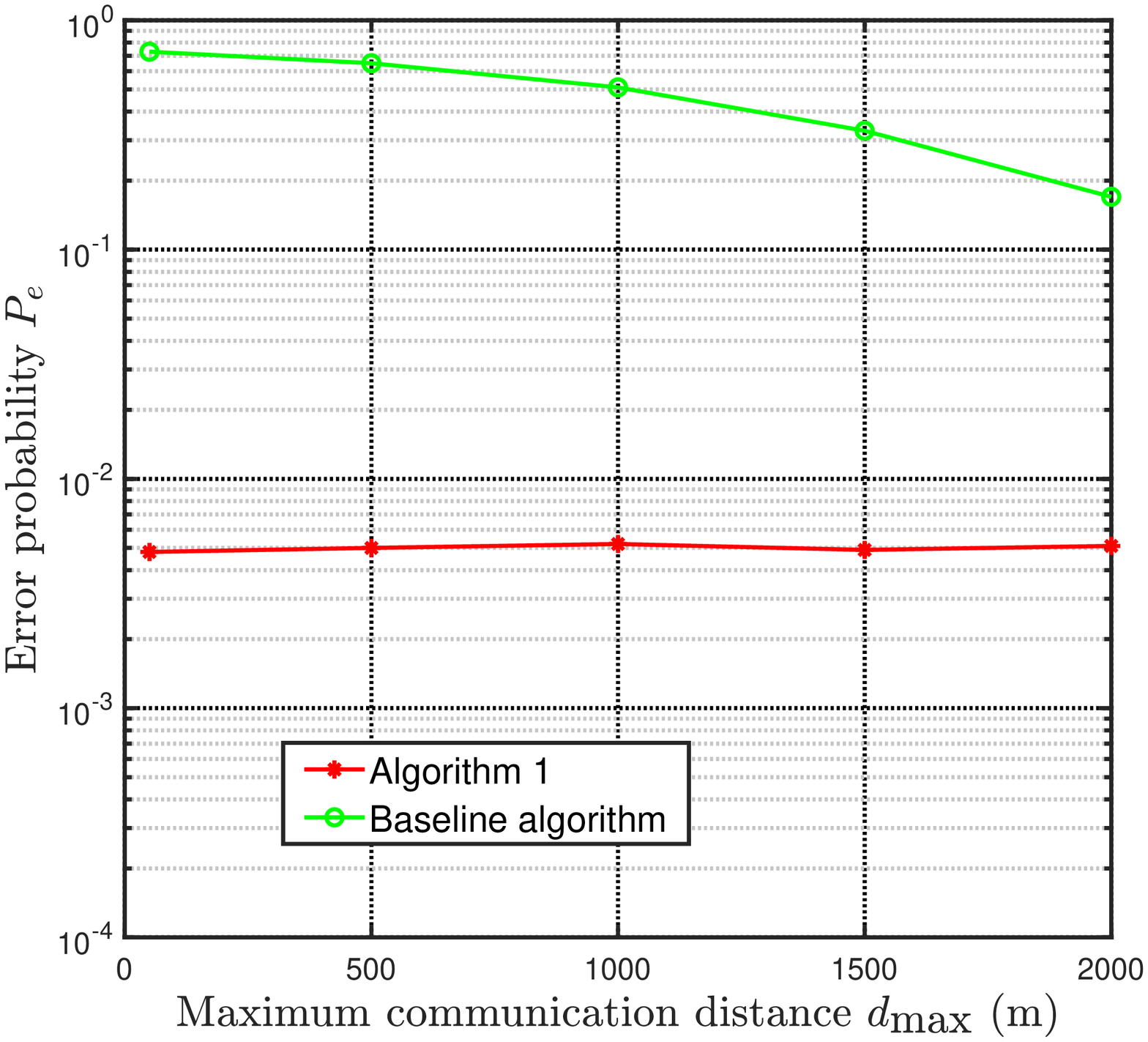}
\caption {The error probability $P_e=P_{\textmd{MD}}+P_{\textmd{FA}}$ versus the maximum communication distance $d_{\textrm{max}}$ for the proposed Algorithm 1 and the baseline algorithm. $K_a=100$, $M=128$.} \label{Fig11}
\end{figure}
In Fig. \ref{Fig11}, we investigate the impact of the maximum communication distance $d_{\textrm{max}}$ \big($d_k \in (0,d_{\textrm{max}})\ \textmd{m}, \forall k$\big) on the total error probability $P_e=P_{\textmd{MD}}+P_{\textmd{FA}}$ for two algorithms. It is observed that with the increment of the maximum communication distance, the error probability of the baseline algorithm remains at large values and begins to decrease until the maximum distance increases to thousands of meters, while the proposed Algorithm 1 performs well offering a stable performance for different maximum communication distances. These results are consistent with the theoretical model above. As the communication distance $d_k$ increases, the Rician factor $\kappa_k$ approaches zero, where the uplink channel $\textbf{h}_k$ in (\ref{channel}) is approximately equal to $\textbf{h}_k^S$. In this case, the Rician fading channel degenerates to the Rayleigh fading channel, which is a subcase of the proposed framework.
However, the LOS propagation is nonnegligible in practical wireless networks with short-distance communications. Therefore, Algorithm 1 has great advantages over a wide range of communication distance.

\begin{figure}[h] \centering
\includegraphics [width=0.5\textwidth] {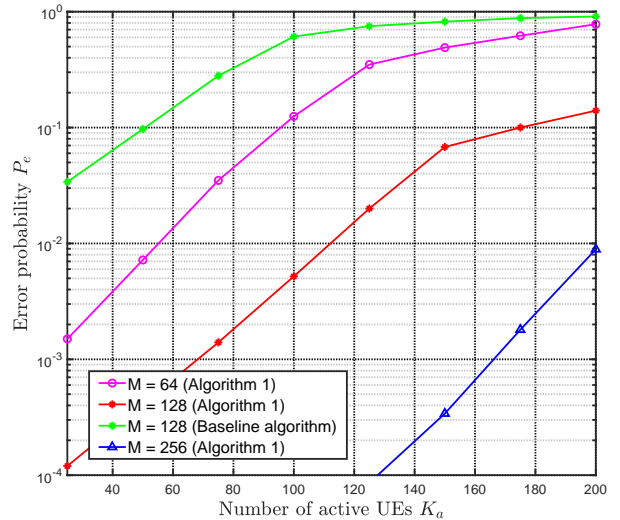}
\caption {The error probability $P_e=P_{\textmd{MD}}+P_{\textmd{FA}}$ versus the numbers of active UEs for different numbers of BS antennas.} \label{Fig7}
\end{figure}
Next, the influence of the number of active UEs $K_a$ on the total error probability $P_e=P_{\textmd{MD}}+P_{\textmd{FA}}$ is evaluated in Fig. \ref{Fig7}. Intuitively, the error probability increases with the increment of $K_a$ rapidly at the beginning and with a diminishing growth rate for a large number of active UEs. This is due to the fact that a larger $K_a$ will cause the outer decoder to include more sub-blocks to check for stitching, resulting in a larger $P_{\textmd{FA}}$. On the other hand, the error probability will improve as the number of BS antennas increases due to more array gains at the BS. It is found that the error probability $P_e$ in the case of $M=256$ is below $10^{-2}$ when there are $200$ active UEs, which is ten times lower than that in the case of $M=64$ or $M=128$. Therefore, the number of active UEs supported by massive unsourced random access systems can be further enhanced by adding the BS antennas while satisfying the requirements of error probability.
Meanwhile, the performance advantage of proposed Algorithm 1 over the compared algorithm is shown when $M\!=\!128$. It can be found that the performance of the compared algorithm in practical Rician fading scenarios deteriorates dramatically due to the inaccurate estimation of codeword activity indicator.

\begin{figure}[h] \centering
\includegraphics [width=0.5\textwidth] {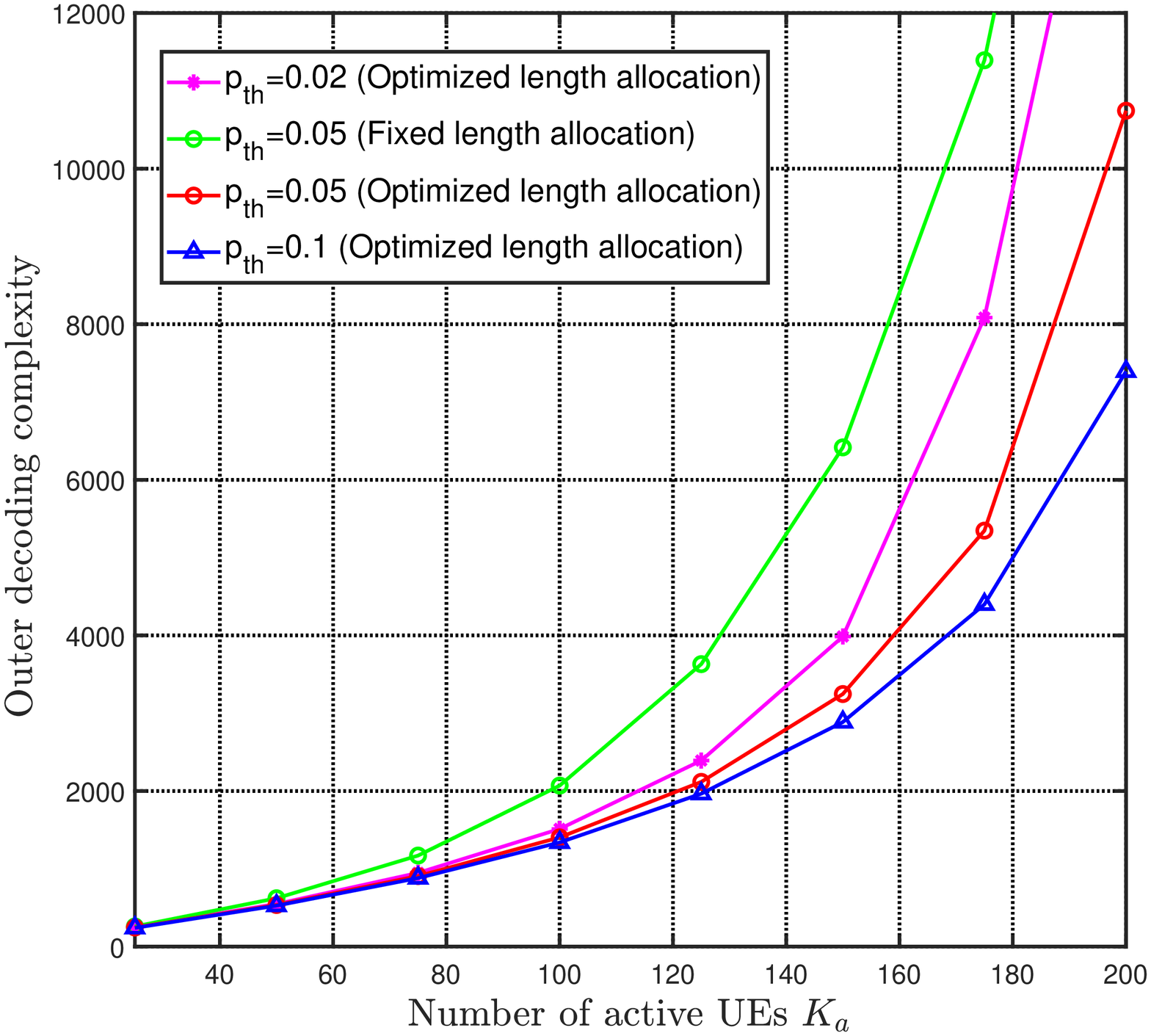}
\caption {The outer decoding complexity $\Xi$ versus the numbers of active UEs for different allocation schemes of parity check bits and different tolerable thresholds of the probability of decoding failure. $J=15$ bits, $L=20$.} \label{Fig89}
\end{figure}
Finally, Fig. \ref{Fig89} reveals the performance advantage of the optimized length allocation and the impact of the number of active UEs on the outer decoding complexity under the optimized length allocation with different tolerable thresholds of the probability of decoding failure.
Given a fixed length allocation of parity check bits $a_l=[7,7,...,7,11,11,11],l\in[2,L]$, the performance gain of the proposed length allocation scheme over the fixed length allocation scheme when $p_{\textmd{th}}=0.05$ is shown. Generally, the optimization algorithm is completely superior to the one adopting a fixed length allocation due to more flexible assignment of parity check bits according to the sequence position of the sub-blocks across the transmission sub-slots.
Meanwhile, it can be seen that the outer decoding complexity $\Xi$ grows as the number of active UEs increases. This is because the width of output list of the inner decoder $K\!=\!\tilde{K}_a+\delta$ becomes larger and the number of nodes that need to be checked in the outer decoding naturally increases, such that the outer decoding is more complicated. Fortunately, it is sensible to relax the threshold of the failure probability for affordable decoding complexity when there are so many active UEs. Therefore, it is necessary to choose a proper $p_{\textmd{th}}$ for striking a balance between service fidelity and system capacity.

In summary, the proposed codeword activity detection algorithm for inner code and the optimized length allocation scheme for outer code have a promising potential of improving the performance of massive unsourced random access in 6G wireless networks.

\section{Conclusion}
In this paper, a general two-layer massive unsourced random access framework, including an inner code component and an outer code component, was designed according to the characteristics of mMTC in 6G wireless networks. Specifically, considering short-distance and short-packet communications, we proposed a codeword activity detection algorithm for the inner code based on the distribution of received signals over Rician fading channels. In particular, the performance of the proposed algorithm was analyzed based on the approximated distribution of estimation error. Moreover, an optimized length allocation algorithm between the lengths of parity check bits and message bits was provided for further improving the efficiency of the outer decoding.
Both theoretical analysis and extensive simulation results confirmed the feasibility and practicability of the proposed massive unsourced random access scheme for 6G wireless networks.

\begin{appendices}
\section{Derivation of the gradient of $q(\tilde{\boldsymbol{\gamma}})$}
Subsequent derivations are based on the following properties of matrix differentiation and transformation:
\begin{eqnarray}
&&\textmd{tr}(\textbf{A}+\textbf{B})=\textmd{tr}(\textbf{A})+\textmd{tr}(\textbf{B}),\\\nonumber
&&\textmd{tr}(\textbf{ABC})=\textmd{tr}(\textbf{BCA})=\textmd{tr}(\textbf{CAB}),\\\nonumber
&&d\textmd{tr}(\textbf{X})=\textmd{tr}(d\textbf{X}),\\\nonumber
&&d(\textbf{UVW})=d(\textbf{U})\textbf{VW}+\textbf{U}d(\textbf{V})\textbf{W}+\textbf{UV}d(\textbf{W}),\\\nonumber
&&d(\textbf{X}^{-1})=-\textbf{X}^{-1}(d\textbf{X})\textbf{X}^{-1}.
\end{eqnarray}
Given $q(\tilde{\boldsymbol{\gamma}})=\frac{1}{M}\textmd{tr}[\tilde{\boldsymbol{\Sigma}}^{-1}\textbf{C}\tilde{\boldsymbol{\Gamma}} \tilde{\textbf{G}}(\textbf{C}\tilde{\boldsymbol{\Gamma}} \tilde{\textbf{G}})^H]-\frac{2}{M}\textmd{tr}(\tilde{\boldsymbol{\Sigma}}^{-1}\textbf{C}\tilde{\boldsymbol{\Gamma}} \tilde{\textbf{G}}\textbf{Y}^H)$ with $\tilde{\boldsymbol{\Sigma}}=\textbf{C}\tilde{\boldsymbol{\Gamma}}\textbf{C}^H+\sigma^2\textbf{I}_{n_0}$, we respectively differentiate its two terms. Ignoring the constant coefficient $\frac{1}{M}$ temporarily, the differentiation of the first term on the right hand side can be derived as (\ref{eqn100}) at the top of the next page,
\begin{figure*}[ht]
\begin{eqnarray}
&&\!\!\!\!\!\!\!\!d\textmd{tr}[(\textbf{C}\tilde{\boldsymbol{\Gamma}}\textbf{C}^H+\sigma^2\textbf{I}_{n_0})^{-1}\textbf{C}\tilde{\boldsymbol{\Gamma}}\tilde{\textbf{G}}(\textbf{C}\tilde{\boldsymbol{\Gamma}} \tilde{\textbf{G}})^H]\nonumber\\
=&&\!\!\!\!\!\!\!\!d\textmd{tr}[\tilde{\textbf{G}}\tilde{\textbf{G}}^H\tilde{\boldsymbol{\Gamma}}\textbf{C}^H (\textbf{C}\tilde{\boldsymbol{\Gamma}}\textbf{C}^H+\sigma^2\textbf{I}_{n_0})^{-1}\textbf{C}\tilde{\boldsymbol{\Gamma}}]\nonumber\\
=&&\!\!\!\!\!\!\!\!\textmd{tr}[d(\tilde{\textbf{G}}\tilde{\textbf{G}}^H\tilde{\boldsymbol{\Gamma}}\textbf{C}^H) (\textbf{C}\tilde{\boldsymbol{\Gamma}}\textbf{C}^H+\sigma^2\textbf{I}_{n_0})^{-1}\textbf{C}\tilde{\boldsymbol{\Gamma}} +\tilde{\textbf{G}}\tilde{\textbf{G}}^H\tilde{\boldsymbol{\Gamma}}\textbf{C}^H d(\textbf{C}\tilde{\boldsymbol{\Gamma}}\textbf{C}^H+\sigma^2\textbf{I}_{n_0})^{-1}\textbf{C}\tilde{\boldsymbol{\Gamma}} +\tilde{\textbf{G}}\tilde{\textbf{G}}^H\tilde{\boldsymbol{\Gamma}}\textbf{C}^H (\textbf{C}\tilde{\boldsymbol{\Gamma}}\textbf{C}^H+\sigma^2\textbf{I}_{n_0})^{-1}d(\textbf{C}\tilde{\boldsymbol{\Gamma}})]\nonumber\\
=&&\!\!\!\!\!\!\!\!\textmd{tr}[\tilde{\textbf{G}}\tilde{\textbf{G}}^Hd\tilde{\boldsymbol{\Gamma}}\textbf{C}^H (\textbf{C}\tilde{\boldsymbol{\Gamma}}\textbf{C}^H+\sigma^2\textbf{I}_{n_0})^{-1}\textbf{C}\tilde{\boldsymbol{\Gamma}}] +\textmd{tr}[\tilde{\textbf{G}}\tilde{\textbf{G}}^H\tilde{\boldsymbol{\Gamma}}\textbf{C}^H (\textbf{C}\tilde{\boldsymbol{\Gamma}}\textbf{C}^H+\sigma^2\textbf{I}_{n_0})^{-1}\textbf{C}d\tilde{\boldsymbol{\Gamma}}]\nonumber \\ &&\!\!\!\!\!\!\!\!-\textmd{tr}[\tilde{\textbf{G}}\tilde{\textbf{G}}^H\tilde{\boldsymbol{\Gamma}}\textbf{C}^H (\textbf{C}\tilde{\boldsymbol{\Gamma}}\textbf{C}^H+\sigma^2\textbf{I}_{n_0})^{-1}\textbf{C}d\tilde{\boldsymbol{\Gamma}}\textbf{C}^H(\textbf{C}\tilde{\boldsymbol{\Gamma}}\textbf{C}^H+\sigma^2\textbf{I}_{n_0})^{-1}\textbf{C}\tilde{\boldsymbol{\Gamma}}]\nonumber\\
=&&\!\!\!\!\!\!\!\!\textmd{tr}[\textbf{C}^H (\textbf{C}\tilde{\boldsymbol{\Gamma}}\textbf{C}^H+\sigma^2\textbf{I}_{n_0})^{-1}\textbf{C}\tilde{\boldsymbol{\Gamma}}\tilde{\textbf{G}}\tilde{\textbf{G}}^Hd\tilde{\boldsymbol{\Gamma}}] +\textmd{tr}[\tilde{\textbf{G}}\tilde{\textbf{G}}^H\tilde{\boldsymbol{\Gamma}}\textbf{C}^H (\textbf{C}\tilde{\boldsymbol{\Gamma}}\textbf{C}^H+\sigma^2\textbf{I}_{n_0})^{-1}\textbf{C}d\tilde{\boldsymbol{\Gamma}}]\nonumber\\ &&\!\!\!\!\!\!\!\!-\textmd{tr}[\textbf{C}^H(\textbf{C}\tilde{\boldsymbol{\Gamma}}\textbf{C}^H+\sigma^2\textbf{I}_{n_0})^{-1}\textbf{C}\tilde{\boldsymbol{\Gamma}}\tilde{\textbf{G}}\tilde{\textbf{G}}^H\tilde{\boldsymbol{\Gamma}}\textbf{C}^H (\textbf{C}\tilde{\boldsymbol{\Gamma}}\textbf{C}^H+\sigma^2\textbf{I}_{n_0})^{-1}\textbf{C}d\tilde{\boldsymbol{\Gamma}}]\nonumber\\
=&&\!\!\!\!\!\!\!\!\textmd{tr}[(\boldsymbol{\Theta}_1-\boldsymbol{\Theta}_2+\boldsymbol{\Theta}_3)d\tilde{\boldsymbol{\Gamma}}],\label{eqn100}
\end{eqnarray}
\hrulefill
\end{figure*}
where $\boldsymbol{\Theta}_1=\textbf{C}^H (\textbf{C}\tilde{\boldsymbol{\Gamma}}\textbf{C}^H+\sigma^2\textbf{I}_{n_0})^{-1}\textbf{C}\tilde{\boldsymbol{\Gamma}}\tilde{\textbf{G}}\tilde{\textbf{G}}^H$, $\boldsymbol{\Theta}_2=\textbf{C}^H(\textbf{C}\tilde{\boldsymbol{\Gamma}}\textbf{C}^H+\sigma^2\textbf{I}_{n_0})^{-1}\textbf{C}\tilde{\boldsymbol{\Gamma}}\tilde{\textbf{G}}\tilde{\textbf{G}}^H\tilde{\boldsymbol{\Gamma}}\textbf{C}^H (\textbf{C}\tilde{\boldsymbol{\Gamma}}\textbf{C}^H+\sigma^2\textbf{I}_{n_0})^{-1}\textbf{C}$ and $\boldsymbol{\Theta}_3=\tilde{\textbf{G}}\tilde{\textbf{G}}^H\tilde{\boldsymbol{\Gamma}}\textbf{C}^H (\textbf{C}\tilde{\boldsymbol{\Gamma}}\textbf{C}^H+\sigma^2\textbf{I}_{n_0})^{-1}\textbf{C}$.

Similarly, the differentiation of the second term of $q(\tilde{\boldsymbol{\gamma}})$ without considering the constant coefficient $-\frac{2}{M}$ temporarily can be computed as (\ref{eqn101}) at the next page,
\begin{figure*}
\begin{eqnarray}
&&\!\!\!\!\!\!\!\!d\textmd{tr}[(\textbf{C}\tilde{\boldsymbol{\Gamma}}\textbf{C}^H+\sigma^2\textbf{I}_{n_0})^{-1}\textbf{C}\tilde{\boldsymbol{\Gamma}}\tilde{\textbf{G}}\textbf{Y}^H]\nonumber\\
=&&\!\!\!\!\!\!\!\!d\textmd{tr}[\tilde{\textbf{G}}\textbf{Y}^H(\textbf{C}\tilde{\boldsymbol{\Gamma}}\textbf{C}^H+\sigma^2\textbf{I}_{n_0})^{-1}\textbf{C}\tilde{\boldsymbol{\Gamma}}]\nonumber\\
=&&\!\!\!\!\!\!\!\!\textmd{tr}[\tilde{\textbf{G}}\textbf{Y}^H d(\textbf{C}\tilde{\boldsymbol{\Gamma}}\textbf{C}^H+\sigma^2\textbf{I}_{n_0})^{-1}\textbf{C}\tilde{\boldsymbol{\Gamma}} +\tilde{\textbf{G}}\textbf{Y}^H(\textbf{C}\tilde{\boldsymbol{\Gamma}}\textbf{C}^H+\sigma^2\textbf{I}_{n_0})^{-1}d(\textbf{C}\tilde{\boldsymbol{\Gamma}})]\nonumber\\
=&&\!\!\!\!\!\!\!\!-\textmd{tr}[\tilde{\textbf{G}}\textbf{Y}^H (\textbf{C}\tilde{\boldsymbol{\Gamma}}\textbf{C}^H+\sigma^2\textbf{I}_{n_0})^{-1}\textbf{C}d\tilde{\boldsymbol{\Gamma}}\textbf{C}^H(\textbf{C}\tilde{\boldsymbol{\Gamma}}\textbf{C}^H+\sigma^2\textbf{I}_{n_0})^{-1}\textbf{C}\tilde{\boldsymbol{\Gamma}}] +\textmd{tr}[\tilde{\textbf{G}}\textbf{Y}^H(\textbf{C}\tilde{\boldsymbol{\Gamma}}\textbf{C}^H+\sigma^2\textbf{I}_{n_0})^{-1}\textbf{C}d\tilde{\boldsymbol{\Gamma}}]\nonumber\\
=&&\!\!\!\!\!\!\!\!-\textmd{tr}[\textbf{C}^H(\textbf{C}\tilde{\boldsymbol{\Gamma}}\textbf{C}^H+\sigma^2\textbf{I}_{n_0})^{-1}\textbf{C}\tilde{\boldsymbol{\Gamma}}\tilde{\textbf{G}}\textbf{Y}^H (\textbf{C}\tilde{\boldsymbol{\Gamma}}\textbf{C}^H+\sigma^2\textbf{I}_{n_0})^{-1}\textbf{C}d\tilde{\boldsymbol{\Gamma}}] +\textmd{tr}[\tilde{\textbf{G}}\textbf{Y}^H(\textbf{C}\tilde{\boldsymbol{\Gamma}}\textbf{C}^H+\sigma^2\textbf{I}_{n_0})^{-1}\textbf{C}d\tilde{\boldsymbol{\Gamma}}]\nonumber\\
=&&\!\!\!\!\!\!\!\!\textmd{tr}[(-\boldsymbol{\Theta}_4+\boldsymbol{\Theta}_5)d\tilde{\boldsymbol{\Gamma}}],\label{eqn101}
\end{eqnarray}
\hrulefill
\end{figure*}
where $\boldsymbol{\Theta}_4=\textbf{C}^H(\textbf{C}\tilde{\boldsymbol{\Gamma}}\textbf{C}^H+\sigma^2\textbf{I}_{n_0})^{-1}\textbf{C}\tilde{\boldsymbol{\Gamma}}\tilde{\textbf{G}}\textbf{Y}^H (\textbf{C}\tilde{\boldsymbol{\Gamma}}\textbf{C}^H+\sigma^2\textbf{I}_{n_0})^{-1}\textbf{C}$ and $\boldsymbol{\Theta}_5=\tilde{\textbf{G}}\textbf{Y}^H(\textbf{C}\tilde{\boldsymbol{\Gamma}}\textbf{C}^H+\sigma^2\textbf{I}_{n_0})^{-1}\textbf{C}$. Therefore, the gradient vector of $q(\tilde{\boldsymbol{\gamma}})$ is given by
\begin{equation}
\triangledown q(\tilde{\boldsymbol{\gamma}})= \textmd{diag}[\frac{1}{M}(\boldsymbol{\Theta}_1-\boldsymbol{\Theta}_2+\boldsymbol{\Theta}_3)^T-\frac{2}{M}(-\boldsymbol{\Theta}_4+\boldsymbol{\Theta}_5)^T].
\end{equation}

\section{Derivation of the Fisher information matrix}
Based on (\ref{log-likeli}), $\log\mathbb{P}(\textbf{Y}|\tilde{\boldsymbol{\gamma}})$ is given by
\begin{eqnarray}
\log\mathbb{P}(\textbf{Y}|\tilde{\boldsymbol{\gamma}})=\!\!\!\!&-&\!\!\!\!M\log|\tilde{\boldsymbol{\Sigma}}|-\textmd{tr}(\tilde{\boldsymbol{\Sigma}}^{-1}\textbf{YY}^H)\nonumber\\
&-&\!\!\!\!\textmd{tr}[\tilde{\boldsymbol{\Sigma}}^{-1}\textbf{C}\tilde{\boldsymbol{\Gamma}} \tilde{\textbf{G}}(\textbf{C}\tilde{\boldsymbol{\Gamma}} \tilde{\textbf{G}})^H]\nonumber\\
&+&\!\!\!\!2\textmd{tr}(\tilde{\boldsymbol{\Sigma}}^{-1}\textbf{C}\tilde{\boldsymbol{\Gamma}} \tilde{\textbf{G}}\textbf{Y}^H).
\end{eqnarray}
The second-order derivative of it with respect to $\tilde{\boldsymbol{\gamma}}_i$ and $\tilde{\boldsymbol{\gamma}}_j$ can be written as
\begin{eqnarray}\label{Fisher}
\frac{\partial^2\log\mathbb{P}(\textbf{Y}|\tilde{\boldsymbol{\gamma}})}{\partial\tilde{\boldsymbol{\gamma}}_i\partial\tilde{\boldsymbol{\gamma}}_j}
=&&\!\!\!\!\!\!\!\!-M\frac{\partial^2\log|\tilde{\boldsymbol{\Sigma}}|}{\partial\tilde{\boldsymbol{\gamma}}_i\partial\tilde{\boldsymbol{\gamma}}_j} -\frac{\partial^2\textmd{tr}(\tilde{\boldsymbol{\Sigma}}^{-1}\textbf{YY}^H)}{\partial\tilde{\boldsymbol{\gamma}}_i\partial\tilde{\boldsymbol{\gamma}}_j}\nonumber\\ &&\!\!\!\!\!\!\!\!-\frac{\partial^2\textmd{tr}[\tilde{\boldsymbol{\Sigma}}^{-1}\textbf{C}\tilde{\boldsymbol{\Gamma}} \tilde{\textbf{G}}(\textbf{C}\tilde{\boldsymbol{\Gamma}}\tilde{\textbf{G}})^H]}{\partial\tilde{\boldsymbol{\gamma}}_i\partial\tilde{\boldsymbol{\gamma}}_j}\nonumber\\ &&\!\!\!\!\!\!\!\!+2\frac{\partial^2\textmd{tr}(\tilde{\boldsymbol{\Sigma}}^{-1}\textbf{C}\tilde{\boldsymbol{\Gamma}} \tilde{\textbf{G}}\textbf{Y}^H)}{\partial\tilde{\boldsymbol{\gamma}}_i\partial\tilde{\boldsymbol{\gamma}}_j}.
\end{eqnarray}
Then, the first term on the right hand side of (\ref{Fisher}) is derived as
\begin{eqnarray}
\Omega_1\!\!\!\!&=&\!\!\!\!\frac{\partial^2\log|\tilde{\boldsymbol{\Sigma}}|}{\partial\tilde{\boldsymbol{\gamma}}_i\partial\tilde{\boldsymbol{\gamma}}_j} =\frac{\partial}{\partial\tilde{\boldsymbol{\gamma}}_j}[\textmd{tr}(\tilde{\boldsymbol{\Sigma}}^{-1}\textbf{c}_i\textbf{c}_i^H)]\nonumber\\
\!\!\!\!&=&\!\!\!\!-\textmd{tr}(\tilde{\boldsymbol{\Sigma}}^{-1}\textbf{c}_j\textbf{c}_j^H\tilde{\boldsymbol{\Sigma}}^{-1}\textbf{c}_i\textbf{c}_i^H),
\end{eqnarray}
while the second term can be calculated as
\begin{eqnarray}
\Omega_2\!\!\!\!&=&\!\!\!\!\frac{\partial^2\textmd{tr}(\tilde{\boldsymbol{\Sigma}}^{-1}\textbf{YY}^H)}{\partial\tilde{\boldsymbol{\gamma}}_i\partial\tilde{\boldsymbol{\gamma}}_j} =\frac{\partial}{\partial\tilde{\boldsymbol{\gamma}}_j}[-\textmd{tr}(\tilde{\boldsymbol{\Sigma}}^{-1}\textbf{c}_i\textbf{c}_i^H\tilde{\boldsymbol{\Sigma}}^{-1}\textbf{YY}^H)]\nonumber\\
\!\!\!\!&=&\!\!\!\!\textmd{tr}(\tilde{\boldsymbol{\Sigma}}^{-1}\textbf{c}_i\textbf{c}_i^H\tilde{\boldsymbol{\Sigma}}^{-1}\textbf{c}_j\textbf{c}_j^H\tilde{\boldsymbol{\Sigma}}^{-1}\textbf{YY}^H)\nonumber\\ &&\!\!\!\!+\textmd{tr}(\tilde{\boldsymbol{\Sigma}}^{-1}\textbf{c}_j\textbf{c}_j^H\tilde{\boldsymbol{\Sigma}}^{-1}\textbf{c}_i\textbf{c}_i^H\tilde{\boldsymbol{\Sigma}}^{-1}\textbf{YY}^H).
\end{eqnarray}
Next, the third term can be computed as (\ref{eqn102}) at the next page,
\begin{figure*}
\begin{eqnarray}
\Omega_3\!\!\!\!&=&\!\!\!\!\frac{\partial^2\textmd{tr}[\tilde{\boldsymbol{\Sigma}}^{-1}\textbf{C}\tilde{\boldsymbol{\Gamma}} \tilde{\textbf{G}}(\textbf{C}\tilde{\boldsymbol{\Gamma}}\tilde{\textbf{G}})^H]}{\partial\tilde{\boldsymbol{\gamma}}_i\partial\tilde{\boldsymbol{\gamma}}_j}\nonumber\\
\!\!\!\!&=&\!\!\!\!\frac{\partial}{\partial\tilde{\boldsymbol{\gamma}}_j}[\textmd{tr}(\tilde{\boldsymbol{\Sigma}}^{-1}\textbf{C}\tilde{\boldsymbol{\Gamma}}\tilde{\textbf{G}}\tilde{\textbf{G}}^H\textbf{C}^H) -\textmd{tr}(\tilde{\boldsymbol{\Sigma}}^{-1}\textbf{C}\tilde{\boldsymbol{\Gamma}}\tilde{\textbf{G}}\tilde{\textbf{G}}^H\tilde{\boldsymbol{\Gamma}}\textbf{C}^H\tilde{\boldsymbol{\Sigma}}^{-1}\textbf{c}_i\textbf{c}_i^H) +\textmd{tr}(\tilde{\boldsymbol{\Sigma}}^{-1}\textbf{C}\tilde{\textbf{G}}\tilde{\textbf{G}}^H\tilde{\boldsymbol{\Gamma}}\textbf{C}^H)]\nonumber\\
\!\!\!\!&=&\!\!\!\!2\textmd{tr}(\tilde{\boldsymbol{\Sigma}}^{-1}\textbf{C}\tilde{\textbf{G}}\tilde{\textbf{G}}^H\textbf{C}^H) -\textmd{tr}(\tilde{\boldsymbol{\Sigma}}^{-1}\textbf{c}_j\textbf{c}_j^H\tilde{\boldsymbol{\Sigma}}^{-1}\textbf{C}\tilde{\boldsymbol{\Gamma}}\tilde{\textbf{G}}\tilde{\textbf{G}}^H\textbf{C}^H) -\textmd{tr}(\tilde{\boldsymbol{\Sigma}}^{-1}\textbf{c}_i\textbf{c}_i^H\tilde{\boldsymbol{\Sigma}}^{-1}\textbf{C}\tilde{\boldsymbol{\Gamma}}\tilde{\textbf{G}}\tilde{\textbf{G}}^H\textbf{C}^H)\nonumber\\
&&\!\!\!\!+\textmd{tr}(\tilde{\boldsymbol{\Sigma}}^{-1}\textbf{c}_i\textbf{c}_i^H\tilde{\boldsymbol{\Sigma}}^{-1}\textbf{c}_j\textbf{c}_j^H\tilde{\boldsymbol{\Sigma}}^{-1}\textbf{C}\tilde{\boldsymbol{\Gamma}}\tilde{\textbf{G}}\tilde{\textbf{G}}^H\tilde{\boldsymbol{\Gamma}}\textbf{C}^H) -\textmd{tr}(\tilde{\boldsymbol{\Sigma}}^{-1}\textbf{c}_i\textbf{c}_i^H\tilde{\boldsymbol{\Sigma}}^{-1}\textbf{C}\tilde{\textbf{G}}\tilde{\textbf{G}}^H\tilde{\boldsymbol{\Gamma}}\textbf{C}^H)\nonumber\\
&&\!\!\!\!+\textmd{tr}(\tilde{\boldsymbol{\Sigma}}^{-1}\textbf{c}_j\textbf{c}_j^H\tilde{\boldsymbol{\Sigma}}^{-1}\textbf{c}_i\textbf{c}_i^H\tilde{\boldsymbol{\Sigma}}^{-1}\textbf{C}\tilde{\boldsymbol{\Gamma}}\tilde{\textbf{G}}\tilde{\textbf{G}}^H\tilde{\boldsymbol{\Gamma}}\textbf{C}^H) -\textmd{tr}(\tilde{\boldsymbol{\Sigma}}^{-1}\textbf{c}_j\textbf{c}_j^H\tilde{\boldsymbol{\Sigma}}^{-1}\textbf{C}\tilde{\textbf{G}}\tilde{\textbf{G}}^H\tilde{\boldsymbol{\Gamma}}\textbf{C}^H),\label{eqn102}
\end{eqnarray}
\hrulefill
\end{figure*}
and finally the last term can be derived as
\begin{eqnarray}\label{lastterm}
\Omega_4\!\!\!\!&=&\!\!\!\!\frac{\partial^2\textmd{tr}(\tilde{\boldsymbol{\Sigma}}^{-1}\textbf{C}\tilde{\boldsymbol{\Gamma}} \tilde{\textbf{G}}\textbf{Y}^H)}{\partial\tilde{\boldsymbol{\gamma}}_i\partial\tilde{\boldsymbol{\gamma}}_j}\nonumber\\
\!\!\!\!&=&\!\!\!\!\frac{\partial}{\partial\tilde{\boldsymbol{\gamma}}_j}[\textmd{tr}(\tilde{\boldsymbol{\Sigma}}^{-1}\textbf{C}\tilde{\textbf{G}}\textbf{Y}^H) -\textmd{tr}(\tilde{\boldsymbol{\Sigma}}^{-1}\textbf{c}_i\textbf{c}_i^H\tilde{\boldsymbol{\Sigma}}^{-1}\textbf{C}\tilde{\boldsymbol{\Gamma}}\tilde{\textbf{G}}\textbf{Y}^H)]\nonumber\\
\!\!\!\!&=&\!\!\!\!\textmd{tr}(\tilde{\boldsymbol{\Sigma}}^{-1}\textbf{c}_i\textbf{c}_i^H\tilde{\boldsymbol{\Sigma}}^{-1}\textbf{c}_j\textbf{c}_j^H\tilde{\boldsymbol{\Sigma}}^{-1}\textbf{C}\tilde{\boldsymbol{\Gamma}}\tilde{\textbf{G}}\textbf{Y}^H)\nonumber\\ &&\!\!\!\!-\textmd{tr}(\tilde{\boldsymbol{\Sigma}}^{-1}\textbf{c}_j\textbf{c}_j^H\tilde{\boldsymbol{\Sigma}}^{-1}\textbf{C}\tilde{\textbf{G}}\textbf{Y}^H)\nonumber\\
&&\!\!\!\!+\textmd{tr}(\tilde{\boldsymbol{\Sigma}}^{-1}\textbf{c}_j\textbf{c}_j^H\tilde{\boldsymbol{\Sigma}}^{-1}\textbf{c}_i\textbf{c}_i^H\tilde{\boldsymbol{\Sigma}}^{-1}\textbf{C}\tilde{\boldsymbol{\Gamma}}\tilde{\textbf{G}}\textbf{Y}^H)\nonumber\\ &&\!\!\!\!-\textmd{tr}(\tilde{\boldsymbol{\Sigma}}^{-1}\textbf{c}_i\textbf{c}_i^H\tilde{\boldsymbol{\Sigma}}^{-1}\textbf{C}\tilde{\textbf{G}}\textbf{Y}^H).
\end{eqnarray}

According to (\ref{Fisher})-(\ref{lastterm}) and the fact that $\mathbb{E}[\textbf{YY}^H]=M\tilde{\boldsymbol{\Sigma}}$ and $\mathbb{E}[\textbf{Y}]=\textbf{C}\tilde{\boldsymbol{\Gamma}}\tilde{\textbf{G}}$, the $(i,j)$-th element of the Fisher Information Matrix can be obtained by taking the expectation with respect to $\textbf{Y}$, i.e.,
\begin{eqnarray}
[\textbf{F}(\tilde{\boldsymbol{\gamma}})]_{ij}\!\!\!\!&=&\!\!\!\!-\mathbb{E}[-M\Omega_1-\Omega_2-\Omega_3+2\Omega_4]\nonumber\\
&=&\!\!\!\!M\Omega_1+\mathbb{E}[\Omega_2]+\Omega_3-2\mathbb{E}[\Omega_4],
\end{eqnarray}
where $\Omega_1$ and $\Omega_3$ are independent of $\textbf{Y}$ and thus regarded as constants. Then due to
\begin{equation}
\mathbb{E}[\Omega_2]=M\textmd{tr}(\tilde{\boldsymbol{\Sigma}}^{-1}\textbf{c}_i\textbf{c}_i^H\tilde{\boldsymbol{\Sigma}}^{-1}\textbf{c}_j\textbf{c}_j^H) +M\textmd{tr}(\tilde{\boldsymbol{\Sigma}}^{-1}\textbf{c}_j\textbf{c}_j^H\tilde{\boldsymbol{\Sigma}}^{-1}\textbf{c}_i\textbf{c}_i^H)
\end{equation}
and
\begin{eqnarray}
\mathbb{E}[\Omega_4]\!\!\!\!&=&\!\!\!\!\textmd{tr}(\tilde{\boldsymbol{\Sigma}}^{-1}\textbf{c}_i\textbf{c}_i^H\tilde{\boldsymbol{\Sigma}}^{-1}\textbf{c}_j\textbf{c}_j^H\tilde{\boldsymbol{\Sigma}}^{-1}\textbf{C}\tilde{\boldsymbol{\Gamma}}\tilde{\textbf{G}}(\textbf{C}\tilde{\boldsymbol{\Gamma}}\tilde{\textbf{G}})^H)\nonumber\\ &&\!\!\!\!-\textmd{tr}(\tilde{\boldsymbol{\Sigma}}^{-1}\textbf{c}_j\textbf{c}_j^H\tilde{\boldsymbol{\Sigma}}^{-1}\textbf{C}\tilde{\textbf{G}}(\textbf{C}\tilde{\boldsymbol{\Gamma}}\tilde{\textbf{G}})^H)\nonumber\\
&&\!\!\!\!+\textmd{tr}(\tilde{\boldsymbol{\Sigma}}^{-1}\textbf{c}_j\textbf{c}_j^H\tilde{\boldsymbol{\Sigma}}^{-1}\textbf{c}_i\textbf{c}_i^H\tilde{\boldsymbol{\Sigma}}^{-1}\textbf{C}\tilde{\boldsymbol{\Gamma}}\tilde{\textbf{G}}(\textbf{C}\tilde{\boldsymbol{\Gamma}}\tilde{\textbf{G}})^H)\nonumber\\ &&\!\!\!\!-\textmd{tr}(\tilde{\boldsymbol{\Sigma}}^{-1}\textbf{c}_i\textbf{c}_i^H\tilde{\boldsymbol{\Sigma}}^{-1}\textbf{C}\tilde{\textbf{G}}(\textbf{C}\tilde{\boldsymbol{\Gamma}}\tilde{\textbf{G}})^H),
\end{eqnarray}
the final form of $[\textbf{F}(\tilde{\boldsymbol{\gamma}})]_{ij}$ after simplification is given by
\begin{eqnarray}\label{finalFisher}
[\textbf{F}(\tilde{\boldsymbol{\gamma}})]_{ij}\!\!\!\!&=&\!\!\!\!M(\textbf{c}_i^H\tilde{\boldsymbol{\Sigma}}^{-1}\textbf{c}_j\textbf{c}_j^H\tilde{\boldsymbol{\Sigma}}^{-1}\textbf{c}_i) +2\textmd{tr}(\tilde{\boldsymbol{\Sigma}}^{-1}\textbf{C}\tilde{\textbf{G}}\tilde{\textbf{G}}^H\textbf{C}^H)\nonumber\\ &&\!\!\!\!-\textbf{c}_j^H\tilde{\boldsymbol{\Sigma}}^{-1}\textbf{C}\tilde{\boldsymbol{\Gamma}}\tilde{\textbf{G}}\tilde{\textbf{G}}^H\textbf{C}^H\tilde{\boldsymbol{\Sigma}}^{-1}\textbf{c}_j\nonumber\\ &&\!\!\!\!-\textbf{c}_i^H\tilde{\boldsymbol{\Sigma}}^{-1}\textbf{C}\tilde{\boldsymbol{\Gamma}}\tilde{\textbf{G}}\tilde{\textbf{G}}^H\textbf{C}^H\tilde{\boldsymbol{\Sigma}}^{-1}\textbf{c}_i\nonumber\\
&&\!\!\!\!-\textbf{c}_i^H\tilde{\boldsymbol{\Sigma}}^{-1}\textbf{c}_j\textbf{c}_j^H\tilde{\boldsymbol{\Sigma}}^{-1}\textbf{C}\tilde{\boldsymbol{\Gamma}}\tilde{\textbf{G}}\tilde{\textbf{G}}^H\tilde{\boldsymbol{\Gamma}}\textbf{C}^H\tilde{\boldsymbol{\Sigma}}^{-1}\textbf{c}_i\nonumber\\ &&\!\!\!\!+\textbf{c}_i^H\tilde{\boldsymbol{\Sigma}}^{-1}\textbf{C}\tilde{\textbf{G}}\tilde{\textbf{G}}^H\tilde{\boldsymbol{\Gamma}}\textbf{C}^H\tilde{\boldsymbol{\Sigma}}^{-1}\textbf{c}_i\nonumber\\
&&\!\!\!\!-\textbf{c}_j^H\tilde{\boldsymbol{\Sigma}}^{-1}\textbf{c}_i\textbf{c}_i^H\tilde{\boldsymbol{\Sigma}}^{-1}\textbf{C}\tilde{\boldsymbol{\Gamma}}\tilde{\textbf{G}}\tilde{\textbf{G}}^H\tilde{\boldsymbol{\Gamma}}\textbf{C}^H\tilde{\boldsymbol{\Sigma}}^{-1}\textbf{c}_j\nonumber\\ &&\!\!\!\!+\textbf{c}_j^H\tilde{\boldsymbol{\Sigma}}^{-1}\textbf{C}\tilde{\textbf{G}}\tilde{\textbf{G}}^H\tilde{\boldsymbol{\Gamma}}\textbf{C}^H\tilde{\boldsymbol{\Sigma}}^{-1}\textbf{c}_j.
\end{eqnarray}
Based on (\ref{finalFisher}), $\textbf{F}(\tilde{\boldsymbol{\gamma}})$ can be written in a matrix form.
\end{appendices}


\begin{thebibliography}{1}
\bibitem{UN0}
F. Tian, X. Chen, L. Liu, and D. W. K. Ng, ``Design of massive unsourced random access over Rician channels," in \emph{Proc. IEEE International Workshop on Signal Processing Advances in Wireless Communications (SPAWC)}, Lucca, Italy, Sept. 2021.

\bibitem{5G1}
X. Chen, D. W. K. Ng, W. Yu, E. G. Larsson, N. Al-Dhahir, and R. Schober, ``Massive access for 5G and beyond," \emph{IEEE J. Sel. Areas Commun.}, vol. 39, no. 3, pp. 615-637, Mar. 2021.

\bibitem{5G2}
N. Ahmed, D. De, and I. Hussain, ``Internet of Things (IoT) for smart precision agriculture and farming in rural areas," \emph{IEEE Internet of Things J.}, vol. 5, no. 6, pp. 4890-4899, Dec. 2018.

\bibitem{6G}
X. Chen, \emph{Massive Access for Cellular Internet of Things Theory and Technique}, Germany: Springer Press, 2019.


\bibitem{GF1}
J. Choi, J. Ding, N. Le, and Z. Ding, ``Grant-free random access in machine-type communication: approaches and challenges," \emph{IEEE Wireless Commun.}, Early access, Sep. 2021.

\bibitem{GF2}
J. Choi, ``On throughput improvement using immediate re-transmission in grant-free random access with massive MIMO," \emph{IEEE Trans. Wireless Commun.}, vol. 19, no. 12, pp. 8341-8350, Sep. 2020.


\bibitem{GF20}
X. Shao, X. Chen, D. W. K. Ng, C. Zhong, and Z. Zhang, ``Cooperative activity detection: Sourced and unsourced massive random access paradigms," \emph{IEEE Trans. Signal Process.}, vol. 68, pp. 6578-6593, 2020.

\bibitem{GF3}
L. Liu and W. Yu, ``Massive connectivity with massive MIMO-Part 1: Device activity detection and channel estimation," \emph{IEEE Trans. Signal Process.}, vol. 66, no. 11, pp. 2933-2946, Jun. 2018.

\bibitem{GF4}
X. Shao, X. Chen, and R. Jia, ``A dimension reduction-based joint activity detection an channel estimation algorithm for massive access," \emph{IEEE Trans. Signal Process.}, vol. 68, pp. 420-435, Dec. 2019.

\bibitem{GF5}
F. Tian and X. Chen, ``Energy-efficient design for massive access in B5G cellular Internet of Things," in \emph{Proc. IEEE Wireless Commun. Netw. Conf. (WCNC)}, pp. 1-6, 2020.

\bibitem{GF51}
M. Ke, Z. Gao, Y. Wu, X. Gao, and R. Schober, ``Compressive sensing-based adaptive active user detection and channel estimation: Massive access meets massive MIMO," \emph{IEEE Trans. Signal Process.}, vol. 68, pp. 764-779, 2020.

\bibitem{GF52}
L. Qiao, J. Zhang, Z. Gao, D. W. K. Ng, M. D. Renzo, and M. -S. Alouini, ``Massive access in media modulation based massive machine-type communications," \emph{IEEE Trans. Wireless Commun.}, vol. 21, no. 1, pp. 339-356, Jan. 2022.

\bibitem{UN1}
Y. Polyanskiy, ``A perspective on massive random access," in \emph{Proc. IEEE Intern. Symp. Inf. Theory (ISIT)}, pp. 2523-2527, 2017.

\bibitem{UN2}
V. K. Amalladinne, J. Chamberland, and K. R. Narayanan, ``A coupled compressive sensing scheme for unsourced multiple access," in \emph{Proc. IEEE Intern. Conf. Acoustics, Speech and Signal Process. (ICASSP)}, pp. 6628-6632, 2018.


\bibitem{UN8}
V. K. Amalladinne, A. K. Pradhan, C. Rush, J. Chamberland, and K. R. Narayanan, ``On approximate message passing for unsourced access with coded compressed sensing," in \emph{Proc. IEEE Intern. Symp. Inf. Theory (ISIT)}, August. 2020.

\bibitem{UN9}
Y. Wu, X. Gao, S. Zhou, W. Yang, Y. Polyanskiy, and G. Caire, ``Massive access for future wireless communication system," \emph{IEEE Wireless Commun.}, vol. 27, no. 4, pp. 148-156, Apr. 2020.

\bibitem{UN10}
A. K. Pradhan, V. K. Amalladinne, K. R. Narayanan, and J. Chamberland, ``Polar coding and random spreading for unsourced multiple access," in \emph{Proc. IEEE Intern. Conf. Commun. (ICC)}, Jul. 2020.


\bibitem{UN4}
A. Fengler, S. Haghighatshoar, P. Jung, and G. Caire, ``Non-Bayesian activity detection, large-scale fading coefficient estimation, and unsourced random access with a massive MIMO receiver," \emph{IEEE Trans. Info. Theory}, vol. 67, no. 5, pp. 2925-2951, Mar. 2021.

\bibitem{UN5}
V. K. Amalladinne, A. K. Pradhan, C. Rush, and J. Chamberland, ``Unsourced random access with coded compressed sensing: Integrating AMP and belief propagation," \emph{arXiv:2010.04364}, 2020.

\bibitem{UN6}
V. Shyianov, F. Bellili, A. Mezghani, and E. Hossain, ``Massive unsourced random access based on uncoupled compressive sensing: Another blessing of massive MIMO," \emph{IEEE J. Sel. Areas Commun.}, vol. 39, no. 3, pp. 820-834, Aug. 2020.




\bibitem{UN11}
S. S. Kowshik, K. Andreev, A. Frolov, and Y. Polyanskiy, ``Energy efficient random access for the quasi-static fading MAC," in \emph{Proc. IEEE Int. Symp. Inf. Theory (ISIT)}, pp. 2768-2772, 2019.

\bibitem{UN12}
A. Decurninge, I. Land, and M. Guillaud, ``Tensor-based modulation for unsourced massive random access," \emph{IEEE Wireless Commun. Lett.}, vol. 10, no. 3, pp. 552-556, Mar. 2021.



\bibitem{Rice1}
J. Zhang, L. Dai, X. Zhang, E. Bjornson, and Z. Wang, ``Achievable rate of Rician large-scale MIMO channels with transceiver hardware impairments," \emph{IEEE Trans. Veh. Technol.}, vol. 65, no. 10, pp. 8800-8806, Oct. 2016.

\bibitem{Rice2}
J. Zhang, L. Dai, Z. He, S. Jin, and X. Li, ``Performance analysis of mixed-ADC massive MIMO systems over Rician fading channels," \emph{IEEE J. Sel. Areas Commun.}, vol. 35, no. 6, pp. 1327-1338, Jun. 2017.




\bibitem{UN13}
X. Xie, Y. Wu, J. Gao, and W. Zhang, ``Massive unsourced random access for massive MIMO correlated channels," in \emph{Proc. IEEE Global Commun. Conf. (GLOBECOM)}, pp. 1-6, 2020.


\bibitem{AD1}
S. Haghighatshoar, P. Jung, and G. Caire, ``Improved scaling law for activity detection in massive MIMO systems," in \emph{Proc. IEEE Intern. Symp. Inf. Theory (ISIT)}, pp. 381-385, 2018.

\bibitem{UN3}
A. Fengler, G. Caire, P. Jung, and S. Haghighatshoar, ``Massive MIMO unsourced random access," \emph{arXiv:1901.00828}, 2019.


\bibitem{CCCP1}
T. D. Quoc and M. Diehl, ``Sequential convex programming methods for solving nonlinear optimization problems with DC constraints," \emph{arXiv:1107.5841}, 2011.

\bibitem{CD1}
P. Tseng, ``Convergence of a block coordinate descent method for nondifferentiable minimization," \emph{J. Optimization Theory and Applications}, vol. 109, no. 3, pp. 475-494, Jun. 2001.

\bibitem{CD2}
J. Sherman and W. J. Morrison, ``Adjustment of an inverse matrix corresponding to a change in one element of a given matrix," \emph{The Annals of Mathematical Statistics}, vol. 21, no. 1, pp. 124-127, 1950.

\bibitem{CCCP2}
R. Horst and N. V. Thoai, ``DC programming: overview," \emph{J. Optimization Theory and Applications}, vol. 103, no. 1, pp. 1-43, Oct. 1999.

\bibitem{SP}
S. M. Kay, \emph{Fundamentals of Statistical Signal Processing: Estimation Theory}. Englewood Cliffs, NJ: Prentice-Hall, 1993.

\bibitem{CVX}
S. Boyd and L. Vandenberghe, \emph{Convex Optimization.} Cambridge, U.K.: Cambridge Univ. Press, 2004.

\bibitem{Rice3}
O. Ozdogan, E. Bjornson, and E. G. Larsson, ``Massive MIMO with spatially correlated Rician fading channels," \emph{IEEE Trans. Commun.}, vol. 67, no. 5, pp. 3234-3250, May. 2019.

\end{thebibliography}
\end{document}